\documentclass[aps,prb,twocolumn,groupedaddress,showpacs,intlimits,amsmath,amssymb,floatfix,citeautoscript,footinbib]{revtex4-1}

\usepackage[final]{graphicx}
\usepackage[hang,center]{subfigure}
\usepackage{amssymb}
\usepackage{amsmath}
\usepackage{xcolor}

\emergencystretch=\maxdimen
\begin{document}
\title{\textcolor{black}{Coupled structural distortions, domains, and control of phase competition in polar SmBaMn$_2$O$_6$}}
\author{Elizabeth A. Nowadnick$^{1,2}$}
\email{E-mail: enowadnick@ucmerced.edu}
\altaffiliation{Current Address: Department of Materials Science and Engineering, University of California: Merced, Merced, CA 95343}
\author{Jiangang He$^{2,3}$}
\author{Craig J. Fennie$^2$}
\affiliation{$^1$Department of Physics, New Jersey Institute of Technology, Newark, NJ 07102}
\affiliation{$^2$School of Applied and Engineering Physics, Cornell University, Ithaca, NY 14853}
\affiliation{$^3$Department of Materials Science and Engineering, Northwestern University, Evanston, IL 60608}
\date{\today}

\begin{abstract}
Materials with coupled or competing order parameters display highly tunable ground states, where subtle perturbations reveal distinct electronic and magnetic phases. These phases generally are underpinned by complex crystal structures, but the role of structural complexity in these phases often is unclear. We use group theoretic methods and first-principles calculations to analyze a set of coupled structural distortions that underlie the polar charge- and orbitally- ordered antiferromagnetic ground state of $A$-site ordered SmBaMn$_2$O$_6$. We show that these distortions play a key role in establishing the ground state and stabilizing a network of domain wall vortices. Furthermore, we show that the crystal structure provides a knob to control competing electronic and magnetic phases at structural domain walls and in epitaxially strained thin films. These results provide new understanding of the complex physics realized across multiple length scales in SmBaMn$_2$O$_6$ and demonstrate a framework for systematic exploration of correlated and structurally complex materials. 

\end{abstract}

\maketitle

\section{Introduction}
Complex materials host multiple coupled or competing structural, electronic, and magnetic order parameters. This complexity manifests itself across multiple length scales. At the microscopic level,  the interplay of multiple degrees of freedom plays a key role in correlated electron phases such as metal-insulator transitions in the nickel and vanadium oxides,~\cite{Imada1998}  high-temperature superconductivity in the copper oxides~\cite{Dagotto2005}, and colossal magnetoresistance  in the manganese oxides~\cite{Tokura2006,Dagotto2013}. At the mesoscale, complex domain structures involving networks of domain wall vortices and antivortices encode the coupling or competition between multiple order parameters~\cite{Seidel2016,Huang2017}. In between these two extremes, individual domain walls can stabilize  states distinct from the bulk, such as conducting domain walls in insulators~\cite{Seidel2009,Meier2012,Oh2015,Ma2015a,Ma2015},  local ferromagnetism at  domain walls in antiferromagnets~\cite{Farokhipoor2014,Hirose2017}, and the modulation of superconductivity at twin walls~\cite{Noad2018}.   Understanding and ultimately controlling coupled and competing degrees of freedom across this range of length scales  remains a challenge. 

\begin{figure}
\includegraphics[width=0.35\textwidth]{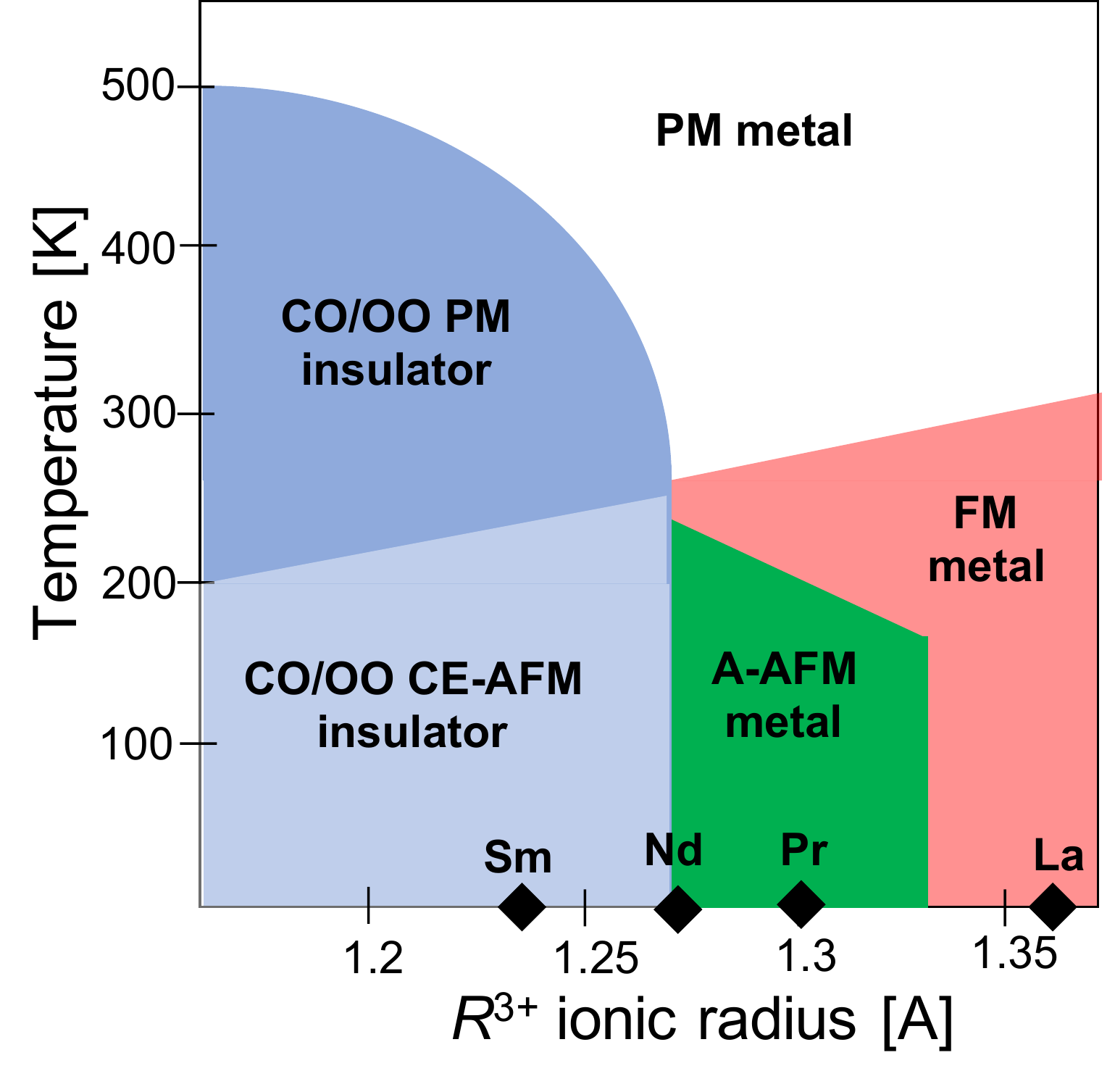}
\caption{\label{fig:phase_diag} {\bf Experimental phase diagram} of the $A$-site ordered $R$BaMn$_2$O$_6$ family ($R$ = rare earth), adapted from Refs.~\onlinecite{Akahoshi2003,Nakajima2003}. Here CO = charge order, OO = orbital order, PM = paramagnetic, FM = ferromagnetic, AFM = antiferromagnetic.}
\end{figure}

The rare earth manganese oxide perovskites provide an ideal context in which to address this challenge, because they host 
several coupled and competing structural, charge order (CO), orbital order (OO), and magnetic phases. 
We concentrate here  on the $A$-site ordered series  $R$BaMn$_2$O$_6$ ($R$ = rare earth); Fig.~\ref{fig:phase_diag} shows the experimental phase diagram of this series.
As the rare earth ionic radius increases, the ground state evolves from a CO/OO CE-type antiferromagnetic (CE-AFM) insulator, to an A-type AFM (A-AFM) metal, and finally to a ferromagnetic (FM) metal~\cite{Akahoshi2003,Nakajima2003,Akahoshi2004_prb}. We focus in particular on SmBaMn$_2$O$_6$, which in addition to being the closest CO/OO/CE-AFM insulating compound to the phase boundary with the A-AFM and FM metallic systems, also possesses a polar crystal structure~\cite{Arima2002,Morikawa2012,Yamada2012,Sagayama2014,Yamaguchi2013}, which  enables additional functionality. 

The coupled CO/OO/CE-AFM state realized in $A$-site ordered SmBaMn$_2$O$_6$ is ubiquitous in half-doped manganites, and has been the subject of extensive investigations~\cite{Tokura2006}. SmBaMn$_2$O$_6$ also has a complex crystal structure characterized by multiple structural distortions that displace the atoms away from their high-symmetry positions. This structural complexity, as well as its interplay with the electronic and magnetic degrees of freedom,  remains much less explored. 
Here, we use group theoretic methods and density functional theory (DFT)+$U$ calculations to analyze the complex ground state crystal structure as well as competing metastable structural phases  of SmBaMn$_2$O$_6$. We find that the interplay of multiple symmetry-allowed couplings between structural distortions is key for stabilizing not only the structural, but also the electronic and magnetic ground states.  
While the role of the  electronic and magnetic degrees of freedom in the CO/OO/CE-AFM state are well known, the contribution of the structural couplings to stabilizing this state has not been appreciated in the past. 

We then show that the crystal structure is the  key ``control knob to turn'' to realize  competing phases and the novel properties associated with them in SmBaMn$_2$O$_6$. We discuss examples of how to take advantage of this structural control knob across a range of length scales, from single domain thin films, to domain walls in bulk systems. 
Taken together, our results provide new insight into the coupled degrees of freedom in SmBaMn$_2$O$_6$, and also show an approach to analyzing complex crystal structures and their impact on the correlated charge, orbital, and spin degrees of freedom, which is applicable to other systems. In particular, we expect that our analysis also applies to the related  $n$=2 Ruddlesden-Popper Pr(Sr$_{1-x}$Ca$_x$)$_2$Mn$_2$O$_7$, which displays an analogous polar CO/OO/CE-AFM ground state~\cite{Tokunaga2006, Ma_Chao2015,Yamauchi2013a}. 

The outline for the rest of this paper is as follows. In Section~\ref{sec:reproducing} we check that DFT+$U$ provides a satisfactory description of the ground state physics of SmBaMn$_2$O$_6$ for our study. In Section~\ref{sec:decomp} we analyze the crystal structure by decomposing it into symmetry distinct distortions, and we understand the relationship between these distortions via a Landau free energy expansion in Section~\ref{sec:free}. We then apply this analysis in Section~\ref{sec:connection}  to reveal the relationship between the structure and the electronic/magnetic order, and in Section~\ref{sec:ferroelectric}  to elucidate the ferroelectric mechanism. We analyze the low-energy phases that may compete with the ground state in Section~\ref{sec:competing}.  In the next sections, we show how the domain structure encodes the coupled and competing degrees of freedom in SmBaMn$_2$O$_6$: Section~\ref{sec:domains} analyzes the ground state domains and enumerates the types of domain walls, Section~\ref{sec:walls}  shows that the walls organize into a network of domain wall vortices and antivortices, and Section~\ref{sec:scenarios} explores scenarios for the competing phases that may be realized at the domain walls. Finally, Section~\ref{sec:strain} shows how epitaxial strain can control the ground state and stabilize a competing FM phase. In  Section~\ref{sec:discussion} we summarize our results.

\section{\label{sec:reproducing}Reproducing the ground state properties}

In correlated insulating systems that have competing metallic phases close in energy, the sense in which  DFT+$U$ captures the relevant underlying physics must be made clear (see Appendix~\ref{app:methods}  for computational methods).  
In particular, it is critical to check that DFT+$U$ captures both  the details of the experimentally observed ground state crystal structure as well as  the observed ground state charge, spin, and orbital order at the same $U$ value. 
We use the  experimentally reported $R$BaMn$_2$O$_6$ phase diagram shown in Fig.~\ref{fig:phase_diag} to guide our choice of $U$ value. We select a $U$ such that the computed ground state reproduces the experimentally reported ground state structural symmetry and magnetic order for several $R$ (see Appendix~\ref{app:u}). We  hold $U$ fixed  for the remainder of this work. 

We next summarize the relevant experimentally known properties of SmBaMn$_2$O$_6$ and check that our calculations reproduce these properties.  
Due to the large size mismatch between Sm and Ba, SmBaMn$_2$O$_6$ can be stabilized in an $A$-site ordered double perovskite form, with Sm and Ba stacking alternately along the $c$ axis.
At high termperature, SmBaMn$_2$O$_6$ crystallizes in the $Cmmm$ space group, and undergoes two structural-CO/OO transitions as the temperature lowers. At $T_\mathrm{1}$ = 380 K, it transitions to a structure with $Pnam$ symmetry, and  at $T_\mathrm{2}$ = 180 K it transition to the polar ground state structure $P2_1am$\cite{Sagayama2014}. 
These phases differ in the stacking of the CO/OO along the $c$ axis: denoting the two phases of the CO as $\alpha$ and $\beta$, there is $\alpha\alpha\beta\beta$ stacking in $Pnam$ and  $\alpha\alpha$ stacking in $P2_1am$. 
Table~\ref{tab:decomp} compares the structural properties of the $P2_1am$ ground state determined from experiment and obtained from DFT+$U$, which show satisfactory agreement. 

\begin{figure}
\includegraphics[width=0.8\linewidth]{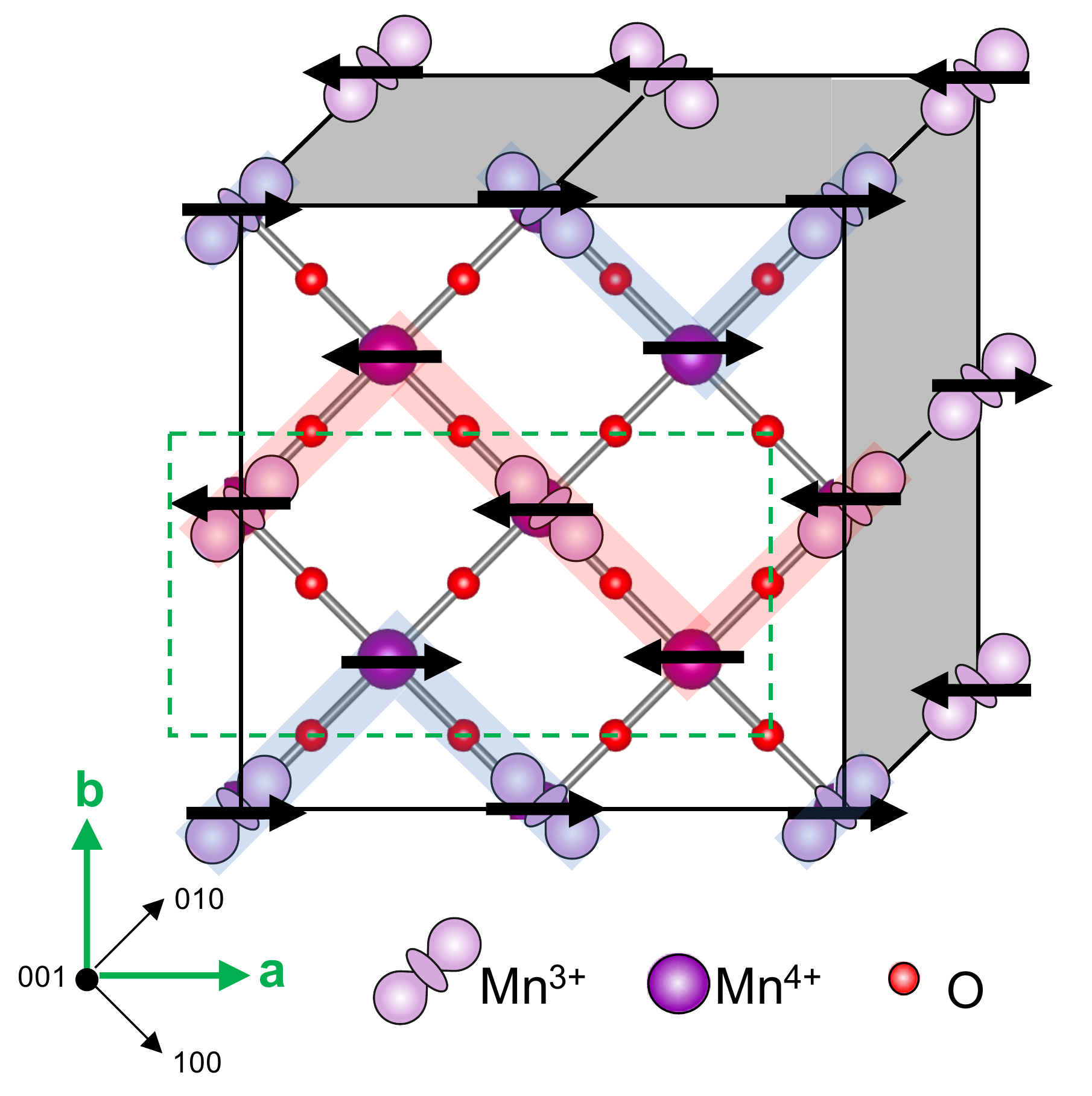}
\caption{ {\bf Coupled ground state charge, orbital, and spin order of SmBaMn$_2$O$_6$.} The Mn$^{3+}$/Mn$^{4+}$ CO forms a checkerboard in the $ab$ plane. The Mn$^{3+}$ sites host a $d_{x^2-r^2}/d_{y^2-r^2}$ OO, which form stripes parallel to the $b$ axis. The spins order in a CE-type AFM pattern, where they form FM-coupled zig-zags in the $ab$ plane (indicated by light blue and red). These zig-zags are AFM-coupled to each other within the $ab$ plane and also along $c$. The crystallographic unit cell is shown by the dashed green box, and the CE-AFM magnetic unit cell is shown by the black box. The green arrows indicate the setting of the orthorhombic axes $a$ and $b$ relative to the tetragonal axes.}
\label{fig:electronic}
\end{figure}

Fig.~\ref{fig:electronic} depicts the experimentally determined electronic and magnetic ground state of SmBaMn$_2$O$_6$. 
The CO typically is understood to arise from a disproportionation of the Mn into Mn$^{3+}$ and Mn$^{4+}$ sites, which form a checkerboard pattern in the $ab$ plane, establishing two sublattices. The  electronic configuration of Mn$^{3+}$ is $t_{2g}^3e_g^1$, so a Jahn-Teller instability yields a $d_{x^2-r^2}/d_{y^2-r^2}$ OO on the Mn$^{3+}$ sublattice~\cite{Wollan1955,Goodenough1955,Tokura2006}. At $T_N = 260$ K, SmBaMn$_2$O$_6$ develops CE-AFM order, which consists of zig-zag chains of FM-coupled spins, where the Mn$^{3+}$ (Mn$^{4+}$) sites sit at the straight sections (corners).  These chains are AFM-coupled to each other within the $ab$ plane and along $c$.  
Our calculations reproduce this coupled CO/OO/CE-AFM state.~\cite{foot1}

\section{\label{sec:decomp} Decomposition of the ground state crystal structure}
 
\begin{figure*}
\centering
\includegraphics[width=0.9\linewidth]{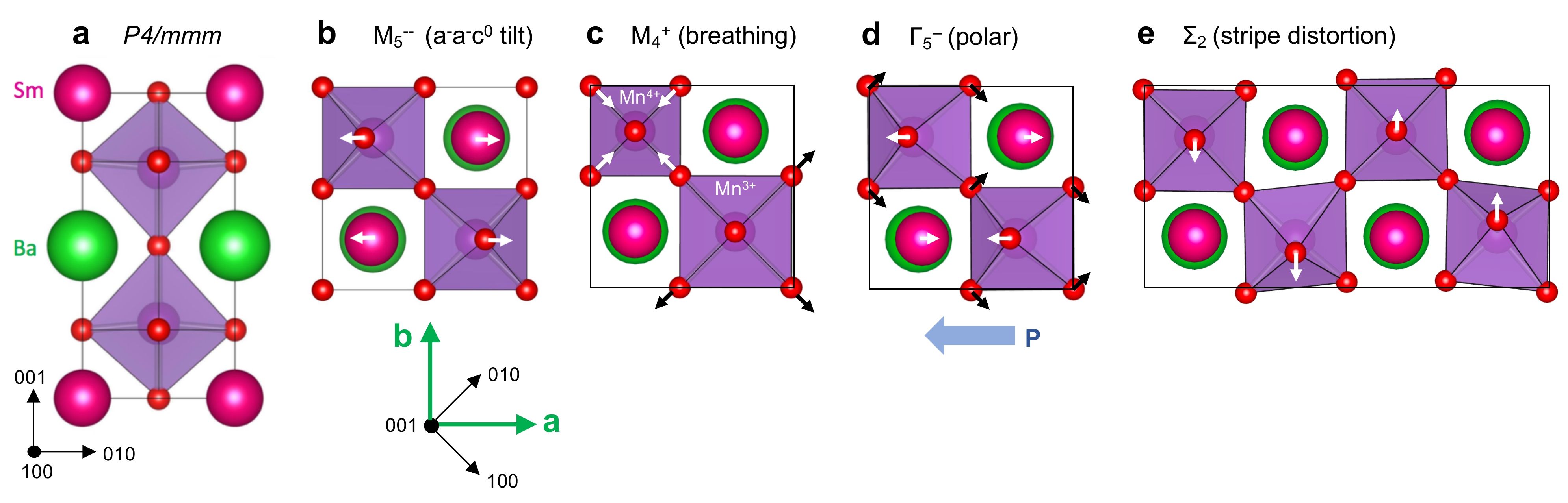}
\caption{ {\bf Structural distortions that contribute to the ground state crystal structure.} (a) High symmetry reference structure $P4/mmm$. The ground state $P2_1am$ structure decomposes into four structural distortions that transform like  irreps of $P4/mmm$: (b) an out-of-phase octahedral tilt that transforms like $M_5^-$, (c) a breathing distortion coupled to the Mn$^{3+}$/Mn$^{4+}$ charge order that transforms like $M_4^+$, (d) a polar mode that transforms like $\Gamma_5^-$, and (e) a set of displacements that form in stripes along $b$ and transform like $\Sigma_2$. In (c) and (d), the distortion amplitudes  are artificially increased for better visualization. For clarity, in (e) only the apical oxygen displacements are highlighted with arrows, although all atoms displace from their high-symmetry positions. The green arrows indicate the setting of the orthorhombic axes $a$ and $b$ relative to the tetragonal axes.}
\label{fig:structure}
\end{figure*}


\begin{table}
\caption{\label{tab:decomp} Decomposition of the $P2_1am$ structure obtained from DFT+$U$ and from  experiment  (Ref.~\onlinecite{Sagayama2014})  into symmetry adapted modes of $P4/mmm$. The amplitudes $\tilde{Q}$ are given for the 40 atom crystallographic unit cell in  \AA. The positive real numbers $a \ne b$ specify the irrep directions.}

\begin{tabular}{c  c  c    c   cl cc}
\hline
\hline
Irrep & Wave- & Direction & Space  && \multicolumn{3}{l}{Amplitude}  \\
& vector& & Group& &&DFT & Expt. \\
\hline
$M_5^-$ & ($\frac{1}{2}$, $\frac{1}{2}$, 0) & (0,-$a$) & $Pbmm$ && $\tilde{Q}_T$ &0.86 & 0.79\\
$M_4^+$ & ($\frac{1}{2}$, $\frac{1}{2}$, 0) & $a$ & $P4/mmm$ &&$\tilde{Q}_b$& 0.20 & 0.19\\
$\Gamma_5^-$& (0,0,0) & (-$a$,-$a$) & $Amm2$ &&$\tilde{Q}_P$& 0.04 & 0.05\\
$\Sigma_2$ & ($\frac{1}{4}$, $\frac{1}{4}$, 0) & ($a$,$b$,0,0) & $P2_1am$ &&$\tilde{Q}_{s}$& 0.91 & 0.71 \\
\hline
\hline
\end{tabular}

\begin{tabular}{c c c}
Lattice parameter [\AA] & DFT & Expt. \\
\hline
$a$  & 11.07 & 11.10 \\
$b$  & 5.53 & 5.54 \\
$c$  & 7.55 & 7.58 \\
\hline
\hline
\end{tabular}
\end{table}

Having confirmed that our computational parameters  reproduce the salient features of the experimental ground state, we now analyze the ground state crystal structure in detail.
The $P2_1am$   structure is highly distorted with respect to the 
  high-symmetry reference structure $P4/mmm$ shown in Fig.~\ref{fig:structure}a.  The $P2_1am$  structure decomposes into four distortions that transform like  irreducible representations (irreps) of $P4/mmm$.  These distortions, shown in Fig.~\ref{fig:structure}b-e, are an out-of-phase ($a^-a^-c^0$ in Glazer notation~\cite{Glazer1972}) octahedral tilting distortion that transforms like  $M_5^-$, a breathing distortion coupled to the Mn$^{3+}$/Mn$^{4+}$  CO that transforms like $M_4^+$,  a polar distortion  that transforms like $\Gamma_5^-$ , and a distortion that  transforms like   $\Sigma_2$. Due to its low symmetry, the $\Sigma_2$ distortion encompasses a complex set of atomic displacements which form stripes parallel to the $b$ axis, we refer to this as the ``stripe distortion" in this work and describe it in more detail in a subsequent section.  
To assess the contributions of these four  distortions to the ground state structure,  Table~\ref{tab:decomp} reports the decomposition of the $P2_1am$ structure, both obtained from experiment (Ref.~\onlinecite{Sagayama2014}) and from DFT+$U$ structural relaxations, into symmetry adapted modes of $P4/mmm$. The $M_5^-$ octahedral tilt and the $\Sigma_2$ stripe distortion have large amplitudes, while the $M_4^+$ breathing and the $\Gamma_5^-$ polar distortions make  smaller contributions.

\section{\label{sec:free}Free energy expansion}
To uncover how the structural distortions described above relate to each other, 
 we perform a Landau free energy expansion about the $P4/mmm$ reference structure. The lowest order terms are:  
\begin{eqnarray}
\label{eq:fe-s}
F_1 &=& \tfrac{1}{2}{\alpha_t}Q_T^2 + \tfrac{1}{4}{\beta_t} Q_T^4 \\ 
&+&\tfrac{1}{2}{\alpha_s}(s_1^2+s_2^2)+ \tfrac{1}{4}{\beta_s}(s_1^2+s_2^2)^2  + \tfrac{1}{2}{\gamma_s}s_1^2s_2^2  \nonumber \\
& + & \tfrac{1}{2}{\alpha_b}Q_{b}^2 + \tfrac{1}{2}{\alpha_p}Q_{P}^2 \nonumber \\
& + & F_{tss} + F_{bss} + F_{tbp} \nonumber
\end{eqnarray}
where
\begin{equation}
\label{eq:tss}
F_{tss} =  \delta_{tss}Q_Ts_1s_2,
\end{equation}
\begin{equation}
\label{eq:css}
F_{bss} =  \delta_{bss}Q_{{b}}(s_1^2-s_2^2),
\end{equation}
and
\begin{equation}
\label{eq:tcp}
F_{tbp} = \delta_{tbp}Q_T Q_{{b}}Q_{P}
\end{equation}
are third-order coupling terms. Here the order parameters $Q_T$, $Q_{b}$, and $Q_{P}$  are the amplitudes of the $M_5^-$ tilt, $M_4^+$ breathing, and $\Gamma_5^-$ polar distortions, respectively.  
The $\Sigma_2$ stripe distortion is described by a two dimensional order parameter ($s_1$,$s_2$), where $s_1$ and $s_2$ give the distortion amplitude on the two sublattices established by the CO/breathing distortion; the total  amplitude is  $Q_{s} = \sqrt{s_1^2 + s_2^2}$.  For simplicity,  Eqs.~\ref{eq:fe-s}-~\ref{eq:tcp} are restricted to one orthorhombic twin of SmBaMn$_2$O$_6$ (treating both twins requires higher dimensional order parameters, as we discuss below). 

\begin{figure*}
\begin{center}
\includegraphics[width=0.75\textwidth]{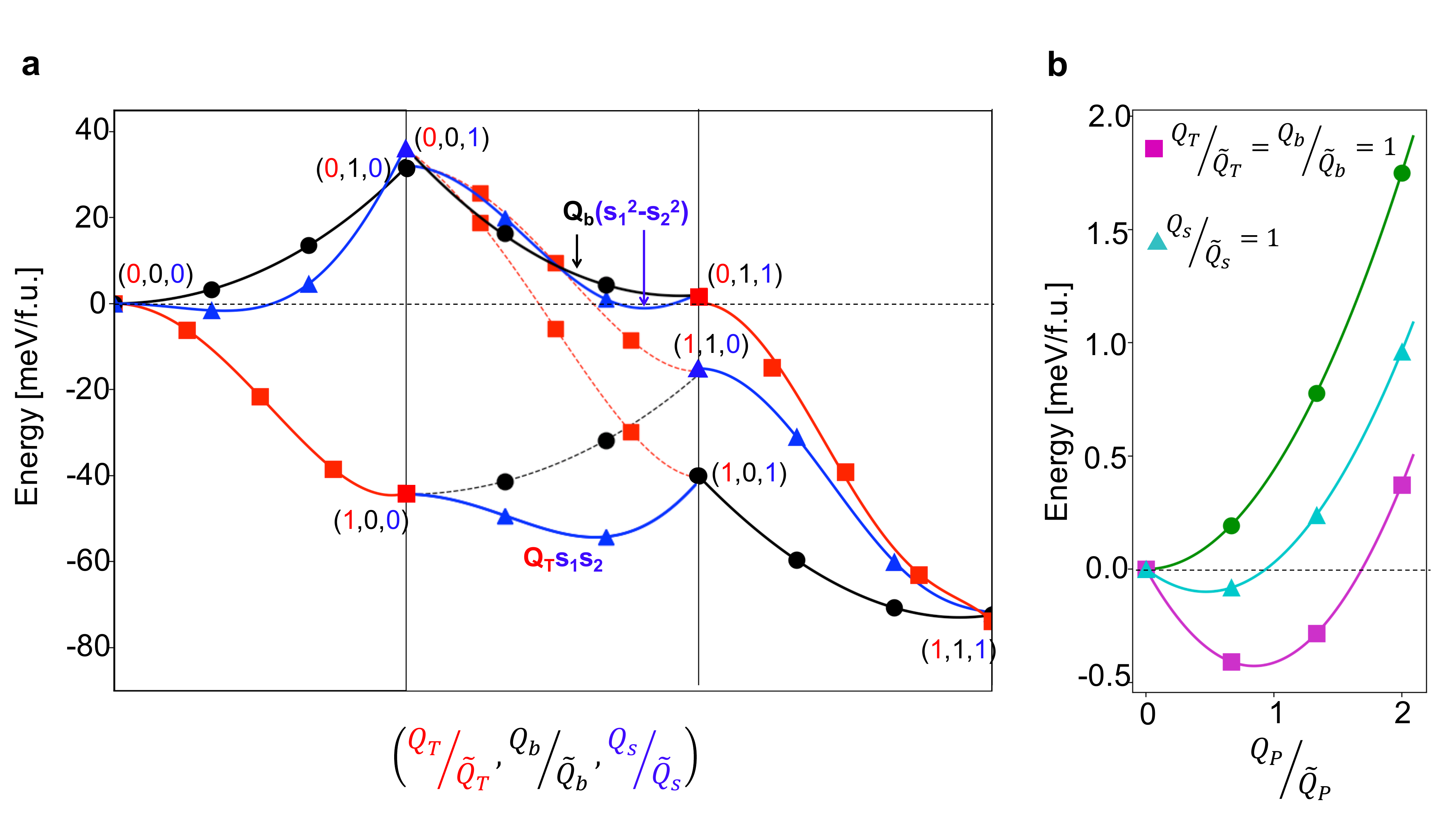}
\caption{\label{fig:surf} {\bf Understanding how  structural distortion couplings establish the  ground state}. The energy surface in the space defined by the $M_5^-$, $M_4^+$, and $\Sigma_2$ distortions is shown in (a). The triplet ($Q_T/\tilde{Q}_T$, $Q_b/\tilde{Q}_{b}$, $Q_s/\tilde{Q}_{s}$) indicates the combination of $M_5^-$, $M_4^+$, and $\Sigma_2$ distortions frozen into $P4/mmm$ along a given branch.  The color and symbol for each curve in each panel  indicates the distortion amplitude that is changing along the curve, so only one distortion is changing along any curve in any panel. In the center panel, the curves discussed in the main text are shown with solid lines, and are labelled by the coupling terms that are active. Here $\tilde{Q}$ are the distortion amplitudes obtained from the DFT+$U$-relaxed $P2_1am$ structure with A-AFM. The stripe distortion order parameter $(s_1,s_2)$ is fixed along the (0.88, 0.47) direction, which is the direction obtained in the DFT+$U$-relaxed $P2_1am$ A-AFM state.  (b) shows energy surfaces obtained by freezing in the $\Gamma_5^-$ distortion alone (green circles), with fixed $Q_s=\tilde{Q}_s$ (cyan triangles), and with fixed $Q_T=\tilde{Q}_T$ and $Q_b=\tilde{Q}_b$ (magenta squares). For all calculations, the lattice parameters are fixed to those of  $P4/mmm$  with A-AFM order, and A-AFM order is imposed (see Appendix~\ref{app:decomp}  for details).}
\end{center}
\end{figure*}

Table~\ref{tab:landau} reports  the coefficients of Eq.~\ref{eq:fe-s},  obtained by freezing combinations of the $M_5^-$, $M_4^+$,  $\Sigma_2$, and $\Gamma_5^-$ distortions into  $P4/mmm$ and fitting the resulting energy surfaces (Fig.~\ref{fig:surf}). For these calculations, we impose A-AFM order and fix the lattice parameters to those of $P4/mmm$ relaxed with A-AFM order. The $P2_1am$ symmetry requires that  $s_1\ne s_2$; we fix the $\Sigma_2$ order parameter direction to be (0.88, 0.47), that is, we freeze in amplitude $Q_s$ where $s_1=0.88Q_s$ and $s_2=0.47Q_s$. This is the direction that occurs in the DFT+$U$-relaxed $P2_1am$ A-AFM state, see Appendices~\ref{app:op} and~\ref{app:stripe} for  details. For clarity, throughout this work we distinguish between \textit{variables} $(s_1,s_2)$ which define an order parameter space, \textit{directions} ($a$,$b$) which are lines through the order parameter space (amplitudes of $a$ and $b$ can vary), and \textit{points} in the order parameter space $(\tilde{s}_1,\tilde{s}_2)$ with fixed amplitude and direction.
  
We first freeze in each distortion individually (Fig.~\ref{fig:surf}a, left panel) and find that $P4/mmm$ is unstable to the $M_5^-$ tilt, very weakly unstable to the $\Sigma_2$ stripe distortion, and stable with respect to the $M_4^+$ breathing distortion. This observation is significant, because it  means that the  couplings in Eqs.~\ref{eq:tss} and~\ref{eq:css} must  induce the $M_4^+$ breathing distortion and the majority of the $\Sigma_2$ stripe distortion amplitude. 

To make this clear, we compute energy surfaces where, in each calculation, we hold one of $Q_T$, $Q_{b}$, and $Q_s$ to a fixed amplitude, and then freeze in one of the remaining two distortions (Fig.~\ref{fig:surf}a, center panel). The lower highlighted branch shows that once the unstable  $M_5^-$ tilt has condensed,  Eq.~\ref{eq:tss}  lowers the energy by inducing a  portion of the $\Sigma_2$ stripe  amplitude. 
The upper highlighted branches show that   Eq.~\ref{eq:css}    strongly lowers the energy (notice the large negative value of $\delta_{bss}$ in Table~\ref{tab:landau}) by  inducing $Q_{b}$ and another portion of the $\Sigma_2$ stripe amplitude.  The right panel of Fig.~\ref{fig:surf}a shows energy surfaces obtained by fixing the amplitudes of two  distortions and freezing in the final one (these again reflect the couplings in Eqs.~\ref{eq:tss}-~\ref{eq:css}).

\begin{table}
\caption{\label{tab:landau} Parameters from fitting the free energy expansion in Eqs.~\ref{eq:fe-s}-~\ref{eq:tcp} to the energy surfaces in Fig.~\ref{fig:surf}. Since we calculate the energy surfaces along a fixed $\Sigma_2$ direction, we omit $\gamma_s$ from Eq.~\ref{eq:fe-s}. 
}
\begin{tabular}{ c| c |c  }
\hline 
$\alpha_t$ -0.31 eV/$\AA^2$ & $\beta_t$ 0.55 eV/$\AA^4$ & $\delta_{tss}$ -0.47 eV/$\AA^3$ \\
$\alpha_s$ -0.05 eV/$\AA^2$  & $\beta_s$ 0.73 eV/$\AA^4$ &  $\delta_{bss}$ -1.48 eV/$\AA^3$\\
 $\alpha_b$ 2.80 eV/$\AA^2$ & & $\delta_{tbp}$ -0.28 eV/$\AA^3$\\
$\alpha_p$ 1.30 eV/$\AA^2$ & \\
\hline
 \end{tabular}
\end{table}

\section{\label{sec:connection}Couplings between the structural, electronic, and magnetic order parameters}
Our observation that $F_{bss}$ induces the  breathing distortion  is significant because this  underlies a direct link between the stripe distortion and the electronic and magnetic degrees of freedom in SmBaMn$_2$O$_6$. To make this clear, we include  electronic and magnetic order parameters in our free energy expansion. Since the electronic CO has the same symmetry as the breathing distortion, its order parameter $Q_{CO}$  couples in the same way to the stripe distortion: 
\begin{equation}
\label{eq:co}
F_{cs} =  \eta_{cs} Q_{CO}(s_1^2-s_2^2).
\end{equation}
Therefore, \textit{by symmetry}, the large amplitude stripe distortion always accompanies the electronic CO, suggesting a far richer electron-lattice coupled state than is typically appreciated in 
 CO models that  focus on the electronic instabilities and associated small amplitude breathing distortions alone. 
Intriguingly, recent  experiments have revealed large-amplitude cation displacements accompanying CO in other manganites~\cite{Savitzky2017,Elbaggari2018}.  

The CE-AFM  also couples directly to the stripe distortion. Two magnetic order parameters define the CE-AFM rstate (one for each sublattice)~\cite{Zhong2000,Ribeiro2016}: $L_\mathrm{CE}=(L_1,L_2)$ and $X_{CE}$, which describe the magnetic ordering on the Mn$^{3+}$ and Mn$^{4+}$ sublattices, respectively. The coupling between $Q_{CO}$ and the CE-AFM order parameters is well known~\cite{Ribeiro2016}:
\begin{equation}
\label{eq:ce}
F_{cL} = \eta_{cL}Q_{CO}(L_\mathrm{1}^2-L_\mathrm{2}^2) + \eta_{cX}Q_{CO}X_{CE}^2. \\
\end{equation}
Combining Eqs.~\ref{eq:co} and ~\ref{eq:ce}, the  coupling between CE-AFM and the stripe distortion is: 
\begin{equation}
\label{eq:ce-s}
F_{sL} = \eta_{sL} (s_1^2-s_2^2)(L_\mathrm{1}^2-L_\mathrm{2}^2) + \eta_{sX}(s_1^2-s_2^2)X_{CE}^2. \\
\end{equation} 
 Eqs.~\ref{eq:co}-~\ref{eq:ce-s} show that the stripe distortion is a  structural control knob to manipulate these electronic and magnetic orders, which we discuss more below.
 
To better understand this important structural distortion,  Table~\ref{tab:sm2}  presents a physically intuitive decomposition of the stripe distortion into several types of atomic displacements (formally, these displacements are all  symmetry adapted modes that transform like $\Sigma_2$, see Appendix~\ref{app:stripe}). 
Starting with the oxygen displacements, a relatively small component  comes from the Jahn-Teller distortion associated with the $d_{x^2-r^2}/d_{y^2-r^2}$ OO on the Mn$^{3+}$ sublattice. The remaining oxygen displacements can be viewed as $a^0a^0c^+$ octahedral rotation-like and $a^-a^-c^0$ octahedral tilt-like distortions on each sublattice that optimize Sm/Ba-O bonding.  The apical oxygen displacements in the Sm layer, which relieve the severe underbonding of the Sm cations, makes by far the largest contribution to the stripe distortion amplitude. Finally,  small displacements of the Sm, Ba, and Mn cations contribute the remaining  amplitude. 
 
\begin{table}
\caption{\label{tab:sm2} Decomposition of the $\Sigma_2$ stripe distortion for both the experimental  (Ref.~\onlinecite{Sagayama2014}) and DFT+$U$-relaxed $P2_1am$ CE-AFM structures  into physically intuitive displacements. The displacements have different amplitudes on the Mn$^{3+}$ and Mn$^{4+}$ sublattices, which are indicated by $\tilde{s}_\mathrm{Mn^{3+}}$ and  $\tilde{s}_\mathrm{Mn^{4+}}$, respectively. The amplitudes are given in \AA~for the 40 atom unit cell.}

\begin{tabular}{c  c     c | c c }
\hline
\hline
Displacement  &  $\tilde{s}_\mathrm{Mn^{3+}}$ &  $\tilde{s}_\mathrm{Mn^{4+}}$ &  $\tilde{s}_\mathrm{Mn^{3+}}$ &  $\tilde{s}_\mathrm{Mn^{4+}}$  \\
Type &  Expt. & Expt. & DFT & DFT\\
\hline
Jahn-Teller (O$_{eq}$) &  0.18 & 0.02 & 0.24 & 0.02\\
\hline
$a^0a^0c^+$ (O$_{eq}$) &  0.34 & 0.04 &0.46 & 0.06\\
\hline
$a^-a^-c^0$ & & & & \\
O$_{ap}$  (Sm layer) & 0.49 & 0.15 &  0.53 & 0.26 \\
O$_{ap}$  (Ba layer) & 0.02 & 0.04 &  0.05 & 0.08\\
O$_{eq}$ & 0.22 & 0.14 & 0.28 & 0.21\\
\hline
Mn  & 0.12 & 0.01 &  0.20 & 0.00\\
Sm  & 0.01 & 0.03 &  0.05 & 0.07\\
Ba  & 0.07 & 0.00 &  0.13 & 0.01\\
\hline
total  & 0.68 & 0.22 &  0.83 & 0.36\\

\end{tabular}
\end{table}

\section{\label{sec:ferroelectric}Ferroelectric mechanism}
We now continue our energy surface analysis to elucidate the ferroelectric mechanism in SmBaMn$_2$O$_6$. Fig.~\ref{fig:surf}b shows that $P4/mmm$ is stable with respect to the $\Gamma_5^-$ polar distortion, which means that a nonlinear coupling to other order parameters must induce the polarization. 
Recently, the concept of ``hybrid improper ferroelectricity" has been introduced in other families of layered perovskite oxides, where a fixed symmetry lowering mechanism (such as layering) removes a subset of symmetries, and then an active set of structural distortions breaks the remaining inversion symmetries and induces a polarization~\cite{Bousquet2008,Benedek2011,Rondinelli2012}. We find that ferroelectricity in SmBaMn$_2$O$_6$ arises from the same crystal chemical idea. First, the Sm/Ba ordering along $c$ removes the inversion centers on the Mn sites. Then, condensation of the coupled $M_5^-$, $M_4^+$, and $\Sigma_2$ distortions  breaks the remaining symmetries,  establishing the $P2_1am$ space group and inducing the polarization.  

All the couplings  in Eqs.~\ref{eq:tss}-~\ref{eq:tcp} are required to establish the polar state. In particular, the interplay of Eq.~\ref{eq:tss} and~\ref{eq:css}  stabilizes the  $\Sigma_2$ order parameter \textit{direction} that establishes the polar $P2_1am$ space group. To specify the symmetry of structures described by multidimensional order parameters, the direction in  order parameter space must be given. As shown in Table~\ref{tab:decomp}, the  ($a$,$b$,0,0) {direction} of  $\Sigma_2$ establishes $P2_1am$, where $a$ and $b$ are real numbers that are not equal to each other (see Appendix~\ref{app:op}). The $\Sigma_2$ order parameter is four dimensional, however, if we work within one orthorhombic twin as we are presently, then we can specify the direction as ($a$,$b$).  

To see how the $(a,b)$ direction is stabilized, note that  Eq.~\ref{eq:tss} is nonzero only if both $s_1$ and $s_2$ are nonzero, and for a fixed  amplitude $Q_s$, is maximal if $s_1=s_2$. This term alone  stabilizes the ($a$,$a$) direction of $\Sigma_2$. In contrast,  Eq.~\ref{eq:css} is nonzero only if $s_1 \ne s_2$, and for a fixed $Q_s$ is maximal if $s_2=0$,  stabilizing the ($a$,0) direction of $\Sigma_2$. As a result, neither term by itself can  establish the ($a$,$b$) direction of $\Sigma_2$, but when they act together this becomes the energetically preferable direction. 

A consequence is that multiple coupling terms induce the polarization. 
The lowest order such term is a trilinear coupling  given in  Eq.~\ref{eq:tcp}. We extend the free energy expansion  to  higher order and find additional coupling terms linear in $Q_P$:
\begin{eqnarray}
\label{eq:fe_sm-p}
F_2 &=&  \zeta_{pts} Q_{P}Q_T(s_1^2-s_2^2) +  \zeta_{pbs} Q_{P}Q_bs_1s_2 \nonumber \\
&+&  \zeta_{pss}Q_{P}s_1s_2(s_1^2-s_2^2).
\end{eqnarray} 

Fig.~\ref{fig:surf}b   shows how multiple coupling terms contribute to inducing the polarization. First holding $Q_T$ and $Q_b$ at  fixed amplitude, we  freeze in $Q_P$ (magenta squares), and find that the energy surface minimum shifts to finite amplitude. This shows that  the trilinear coupling in Eq.~\ref{eq:tcp} induces $Q_P$. Next, holding $Q_s$ fixed and again freezing in $Q_P$  (cyan triangles),  the minimum again shifts to a smaller (but nonzero) amplitude. This shift is smaller because this energy surface reflects the last term  in Eq.~\ref{eq:fe_sm-p}, which is at fifth order. Freezing in different distortion combinations and performing analogous calculations  shows the contributions of the other terms in Eq.~\ref{eq:fe_sm-p}. 

 Since the $M_5^-$, $M_4^+$, and $\Sigma_2$  distortions  act together  to establish the polar state,   SmBaMn$_2$O$_6$ has a  hybrid improper  ferroelectric mechanism. 
Previous work described SmBaMn$_2$O$_6$ as an improper ferroelectric~\cite{Yamaguchi2013}, which does not take into account the combined action of $Q_T$ and $Q_{b}$ in establishing the ($a$,$b$) direction of $\Sigma_2$.

\section{\label{sec:competing}Competing phases}

Our symmetry-based approach provides a natural framework with which to systematically investigate phases that may compete with the bulk ground state of SmBaMn$_2$O$_6$. First, we identify metastable structural phases by exploring the energy landscape in the space defined by the $M_5^-$ and the $\Sigma_2$ distortions, since these distortions  have the largest amplitudes in the ground state structure. Formally, this means that we enumerate all isotropy subgroups generated by distinct directions of the $M_5^-$ and $\Sigma_2$ irreps. Then,  we use DFT+$U$ to investigate the structural, electronic, and magnetic properties of SmBaMn$_2$O$_6$ when the  symmetry is constrained to each of these space groups. We explore all phases with both FM and A-AFM orders imposed, since both  occur in the bulk $R$BaMn$_2$O$_6$ experimental phase diagram. These magnetic orders compete with $Q_b$ and  the stripe distortion (if $s_1 \ne s_2$) via biquadratic couplings:
\begin{eqnarray}
F_{Mb}&=&\eta_{Mb}Q_b^2M^2 + \eta_{Ab}Q_b^2L_A^2 \\ \nonumber
F_{Ms}&=&\eta_{Ms}(s_1^2-s_2^2)M^2 + \eta_{As}(s_1^2-s_2)^2L_A^2
\end{eqnarray}
where $M$ and $L_A$ are the FM and A-AFM order parameters, respectively.

\begin{table}
\caption{\label{tab:m5-sm2} Isotropy subgroups established by selected distinct directions of the $M_5^-$ and $\Sigma_2$ irreps. Energies are obtained from DFT+$U$ structural relaxations with the symmetry constrained to each space group and FM or A-AFM order imposed, and are given in  meV per formula unit (f.u.). The right two columns report the ratio $c/a$ where $c/a=2$ corresponds to zero tetragonal distortion (for orthorhombic space groups, we average the $a$ and $b$ axes to compute these ratios). }

\begin{tabular}{c | c | c | c | c | c | c}
\hline
\hline
$M_5^-$ & $\Sigma_2$ & Space  & Energy  & Energy & $c/a$ & $c/a$\\
& &Group & FM & A-AFM & FM & A-AFM\\
\hline
- & - & $P4/mmm$ & 85.25 & 94.52 & 1.99 &  1.94\\
\hline
 ($a$,0) &- & $Pmam$ & 39.57 & 46.09 & 1.98 & 1.93\\
 ($a$,$a$)&-  & $Cmmm$  & 10.80 & 16.11 & 1.97 & 1.93\\
\hline
 -& ($a$,0,0,0) & $Pmam$ & 83.50 & 74.64 & 1.99 & 1.94\\
(0,$a$) & ($a$,$a$,0,0) & $Pbam$ & 29.13 & 18.52 & 1.97 & 1.93\\
(0,-$a$)&($a$,$b$,0,0) & $P2_1am$ & - & 4.39 & - & 1.93 \\
\hline
& & & & CE-AFM & & CE-AFM \\
(0,-$a$)&($a$,$b$,0,0) & $P2_1am$ & - & 0.0 & - & 1.93 \\
\hline
\end{tabular}
\end{table}

We start with the two-dimensional $M_5^-$ order parameter, which has three symmetry-distinct directions: ($a$,0), ($a$,$a$), and ($a$,$b$). These define space groups $Pmam$, $Cmmm$, and $P2/m$, respectively. The ($a$,0) direction of 
 $M_5^-$ is found in the ground state structure, while  the $(a,a)$ direction gives  the experimentally observed  high-temperature phase ($Cmmm$)~\cite{Sagayama2014}. The difference between these phases is the octahedral tilt axis: the octahedra tilt about the tetragonal [110] axis in $Pmam$, and about [100] in $Cmmm$. 
We next perform DFT+$U$ structural relaxations with the symmetry constrained to $Pmam$ and $Cmmm$ and separately impose both FM and A-AFM orders; the resulting total energies are shown in Table~\ref{tab:m5-sm2}. For both magnetic configurations, the $Cmmm$ structure is lower energy than $Pmam$. In fact, the $Cmmm$ FM phase is only 10.8 meV/f.u. higher in energy than the $P2_1am$ CE-AFM ground state. 

There are many distinct directions and corresponding isotropy subgroups of the $\Sigma_2$ order parameter, since it is four dimensional. Table~\ref{tab:m5-sm2} reports the total energies obtained from  structural relaxations with the symmetry constrained to several of these subgroups~\cite{foot1}. Most directions of $\Sigma_2$, such as the ($a$,0,0,0) direction shown in Table~\ref{tab:m5-sm2}, only slightly lower the energy relative to the high-symmetry reference $P4/mmm$. The lowest energy structural phase above the ground state is $Pbam$, established by the ($a$,$a$,0,0) direction of $\Sigma_2$. $P2_1am$ with A-AFM order is only 4.4 meV/f.u. higher energy than with CE-AFM order (with FM order, it relaxes to the higher-symmetry $Pbam$).  
The competing phases are all centrosymmetric and metallic with the exception of the $P2_1am$ A-AFM phase, which has a small insulating gap.~\cite{foot1} In the next sections, we  discuss how these competing phases may stabilize in SmBaMn$_2$O$_6$. 

\section{\label{sec:domains}Coupled structural domains and domain walls}

\begin{figure*}
\begin{center}
\includegraphics[width=0.9\textwidth]{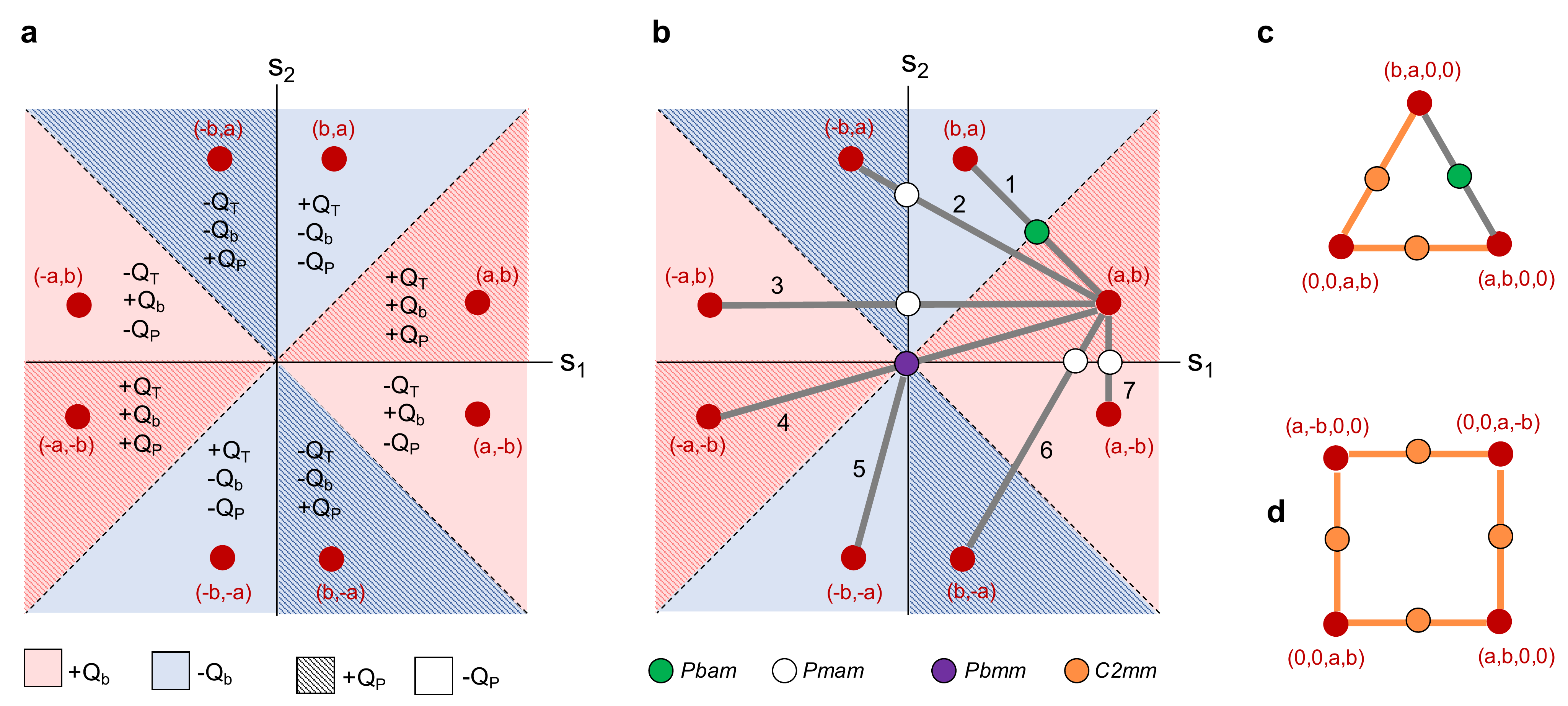}
\caption{\label{fig:switching} {\bf Visualizing coupled structural domains and revealing competing phases.} (a) Structural domains within one orthorhombic twin of SmBaMn$_2$O$_6$.  The domains (red dots) are  points within the two-dimensional space  defined by the $\Sigma_2$  stripe  order parameter directions ($s_1$,$s_2$) for that twin. The  domains in the top right and lower left quadrants have octahedral tilt $Q_T$, while the domains in the other two quadrants have tilt $-Q_T$. The pink (blue) regions denote regions with CO/breathing distortion $+Q_b$ ($-Q_b$), and the hatched (solid) regions indicate regions with polarization $Q_{P}$ (-$Q_{P}$). (b) Paths between domains, indicated by grey lines, with the barrier structures labelled by colored circles: $Pbam$ (green), $Pmam$ (white), and $Pbmm$ (purple).  Extending to the four-dimensional order parameter space defined by the full $\Sigma_2$ order parameter ($s_1$,$s_2$,$s_3$,$s_4$), we can construct closed paths that represent (b) 3-fold and (c) 4-fold domain wall vortices. Here the thick orange lines indicate paths between domains in opposite orthorhombic twins (twin walls), the orange circle is the $C2mm$ barrier for this path.}
\end{center}
\end{figure*}

This section  shows that the domain structure of SmBaMn$_2$O$_6$ encodes the coupling and competition of order parameters  analyzed in previous sections. We first organize the bulk domains, and then explore the domain walls.   

The $P2_1am$  ground state has sixteen structural domains established by the different settings of the ($a$,$b$,0,0) direction of $\Sigma_2$ (listed in Appendix~\ref{app:energy}). These domains are divided between two orthorhombic twins. The eight domains within one  twin are represented as points in the two-dimensional order parameter space ($s_1$,$s_2$) that defines the $\Sigma_2$ distortion in that twin (Fig.~\ref{fig:switching}a). 
Once the  $\Sigma_2$ order parameter is chosen for a given domain, the  couplings in Eqs.~\ref{eq:tss}-~\ref{eq:tcp}  determine the signs of the other structural order parameters in that domain. 
The domains in each quadrant of  Fig.~\ref{fig:switching}a have the same octahedral tilt  ($\pm Q_T$), the domains in the pink (blue) regions have the same CO/breathing distortion $Q_b$ (-$Q_b$), and the domains in the hatched (solid) regions have the same polarization $Q_P$ (-$Q_P$). Thus these structural domains directly visualize the couplings between structural distortions.

There are several distinct types of domain walls that separate the domains  in Fig.~\ref{fig:switching}a. Within order parameter space, domain walls are represented as paths between the domains. These paths can be thought of as  a generalization of intrinsic ferroelectric switching paths, which connect domains with different polarization directions, to other domain types~\cite{Nowadnick2016}. To enumerate the different domain walls  in SmBaMn$_2$O$_6$, we consider all paths that connect the  ($a$,$b$) domain to the other domains (thick grey lines in Fig.~\ref{fig:switching}b). 
The crystal structure evolves along each path, and the highest energy structure gives the energy barrier for that path. Since SmBaMn$_2$O$_6$ is described by several multidimensional order parameters,  there can be more than one type of path between a given pair of domains (in the language of ferroelectric switching, these are one-step versus two-step paths). These paths pass through different sequences of structures and have different energy barriers. While paths through order parameter space  cannot provide quantitative information on domain wall properties (because they rely on bulk calculations), they provide an efficient strategy to organize the many domain wall types.

  All paths in Fig.~\ref{fig:switching}b connect different  $\Sigma_2$ domains, so they all describe stripe distortion domain walls. Paths that cross the $s_1$ and $s_2$ axes also are octahedral tilt antiphase walls, paths that connect domains in blue and pink regions are CO/breathing domain walls, and paths that connect domains in hatched and solid regions are 180$^\circ$ polar walls. 
For example, path 1 in  Fig.~\ref{fig:switching}b describes a CO and  polar domain wall. The   barrier structure for this path is $Pbam$. Paths 2, 3, 6, and 7  all describe octahedral tilt antiphase walls and have barrier structure $Pmam$ (the ($a$,0,0,0) direction of $\Sigma_2$). Paths 2 and 6 (3 and 7) also are CO (polar) walls. Along paths 4 and 5, both $s_1$ and $s_2$ reverse sign, so the barrier structure is $Pbmm$ (the (0,$a$) direction of $M_5^-$). These barrier structures are the competing phases discussed above, so Table~\ref{tab:m5-sm2} reports their energies. 
While the exact barrier energy depends on the magnetic state,  irrespective of the magnetic states  in Table~\ref{tab:m5-sm2}, the energy ordering  from lowest to highest is: $Pbam$ (path 1), $Pbmm$ (paths 4 and 5), $Pmam$ (paths 2, 3, 6, 7). This ordering reflects  that turning off the small amplitude $Q_b$ (path 1) costs less energy than turning off the large amplitude octahedral tilt $Q_T$ (paths 2, 3, 6, 7).

In addition to the paths in Fig.~\ref{fig:switching}b that all stay within one orthorhombic twin, there are paths that connect domains in opposite twins  
 (twin walls). Describing these paths requires the  four dimensional $\Sigma_2$ order parameter ($s_1$,$s_2$,$s_3$,$s_4$);  $s_1$ and $s_2$ are nonzero in one twin, and $s_3$ and $s_4$ are nonzero in the other (see Appendix~\ref{app:stripe}). 
 As an example, we consider the path between the  ($a$,$b$,0,0) and (0,0,$a$,$b$) domains. The barrier structure along this path is the $(a,b,a,b)$ direction of $\Sigma_2$, with symmetry $C2mm$ (all paths between domains in opposite twins have the same  $C2mm$ barrier). The $C2mm$ structure relaxes to  $Cmmm$  (listed in Table~\ref{tab:m5-sm2}) after DFT+$U$ structural relaxations, which has much lower energy than the barriers for the paths in Fig.~\ref{fig:switching}b. 
 
 \section{\label{sec:walls}Domain wall vortices and antivortices}
The observation that the $Cmmm$ barrier is much lower than the other barriers has significant implications for the stability of the different domain wall types. The paths in Fig.~\ref{fig:switching}b are one-step paths, which means that they pass directly from one domain to the other via  one barrier structure. An alternative is a two-step path, where the path passes through a domain in the other orthorhombic twin. For example,  the two-step alternative to path 3 in Fig.~\ref{fig:switching}b is ($a$,$b$,0,0) $\rightarrow$ (0,0,$a$,$b$) $\rightarrow$ (-$a$,$b$,0,0). The barrier for the first and second steps of this path are the  ($a$,$b$,$a$,$b$) and (-$a$,$b$,$a$,$b$) directions of $\Sigma_2$, respectively, which are just different $C2mm$ domains  (which relax to  $Cmmm$).  Therefore, for every path in Fig.~\ref{fig:switching}b, there is an alternative two-step path with barrier $Cmmm$. For a given pair of domains, comparing the one-step barrier to twice the two-step barrier indicates  the lowest energy path. To make this comparison, for each structure we choose the  lowest energy magnetic state from Table~\ref{tab:m5-sm2}, so the barrier energies (in meV/f.u.) are 10.8 ($Cmmm$), 18.5 ($Pbam$), 39.6 ($Pbmm$), and 74.6 ($Pmam$). This  implies that path 1 is the only one-step path that is lower energy than the two-step alternative, so it is energetically favorable for all other paths in Fig.~\ref{fig:switching}b to decay into the two-step path (pairs of twin walls). This does not depend on our particular choice of the barrier magnetic state, for example choosing the barriers to be all A-AFM or all FM leads to the same conclusion.

Structural domain walls terminate at the edge of the sample or they merge with other  walls at domain wall vortices which appear a range of complex materials~\cite{Huang2017}. Within order parameter space, closed paths that start and end at the same domain and are traversed (counter)clockwise represent domain wall (anti)vortices. The number of domains that a closed path passes through gives the order of the vortex. Fig.~\ref{fig:switching}c-d show the domain wall vortices that we find  to be energetically favorable in SmBaMn$_2$O$_6$. Two types of walls are stable against decaying into lower energy walls: path 1 in Fig.~\ref{fig:switching}b (CO/polar wall, barrier $Pbam$) and the path between orthorhombic twins (barrier $Cmmm$). Starting at the ($a$,$b$,0,0) domain, we  construct a closed path  ($a$,$b$,0,0)$\rightarrow$ ($b$,$a$,0,0) $\rightarrow$ (0,0,$a$,$b$) $\rightarrow$ ($a$,$b$,0,0), shown in Fig.~\ref{fig:switching}c. This path describes a 3-fold vortex where one CO/polar wall and two twin walls merge. We also can construct an alternative closed path ($a$,$b$,0,0)$\rightarrow$ (0,0,$a$,$-b$) $\rightarrow$ ($a$,$-b$,0,0) $\rightarrow$ (0,0,$a$,$b$) $\rightarrow$ ($a$,$b$,0,0)  shown in Fig.~\ref{fig:switching}d, which describes a 4-fold vortex where four twin walls meet. This indicates that a domain pattern characterized by a network of  3-fold and 4-fold domain wall vortices is energetically favorable in SmBaMn$_2$O$_6$. This type of 3-fold/4-fold vortex domain structure has been observed in Pr(Sr$_{1-x}$Ca$_x$)$_2$Mn$_2$O$_7$~\cite{Ma2015,Huang2017}.

\section{\label{sec:scenarios}Scenarios for electronic and magnetic domain wall states}
Since domain walls locally modulate the crystal structure, they can  reveal competing phases that are not present in the bulk domains. 
Here, we  speculate about electronic and magnetic states that  may stabilize at SmBaMn$_2$O$_6$  domain walls. The  barrier structures discussed above all host metallic FM or A-AFM states, so if the domain walls are sufficiently wide to realize a bulk-like structure in the middle, these states may naturally stabilize. Furthermore, an implication of the coupling between the stripe distortion and  CE-AFM (Eq.~\ref{eq:ce-s}) is that all structural  walls (with the exception of path 7 in Fig.~\ref{fig:switching}b) are also CE-AFM magnetic  walls. The interruption of the CE-AFM order at the  walls also may allow competing FM and A-AFM states to arise. \textcolor{black}{Interestingly, recent work on manganite strips found that FM metallic edge states form at the edges of the strips, where the CE-AFM order is interrupted.}~\cite{Du2015,Li2016}
While mechanical boundary conditions and local structural relaxations not considered here influence domain wall states, these scenarios show how the naturally occurring structural modulation at domain walls may stabilize competing phases. 

\section{\label{sec:strain}Phase control with epitaxial strain}
As a second example of how structural modulation can control the electronic/magnetic state, we investigate how epitaxial strain impacts single domain SmBaMn$_2$O$_6$ thin films.  Comparing the $c/a$ ratio for the \textit{bulk} ground state and  competing phases  in Table~\ref{tab:m5-sm2}, we find that   
 regardless of the structural symmetry,  $c/a\approx1.98$  with  FM order and $c/a\approx1.94$ for  CE- and A-AFM orders. Thus,  $c/a$  provides a knob to tune between FM and AFM states.  The application of epitaxial strain in thin films is a well-known way to control $c/a$: compressive strain applied in the $ab$ plane shortens $a$ and $b$ while $c$ lengthens  to maintain an approximately constant volume.  
This is different from chemical substitution, where  all lattice parameters change at approximately the same rate (Appendix~\ref{app:phase}). 

\begin{figure}
\includegraphics[width=0.48\textwidth]{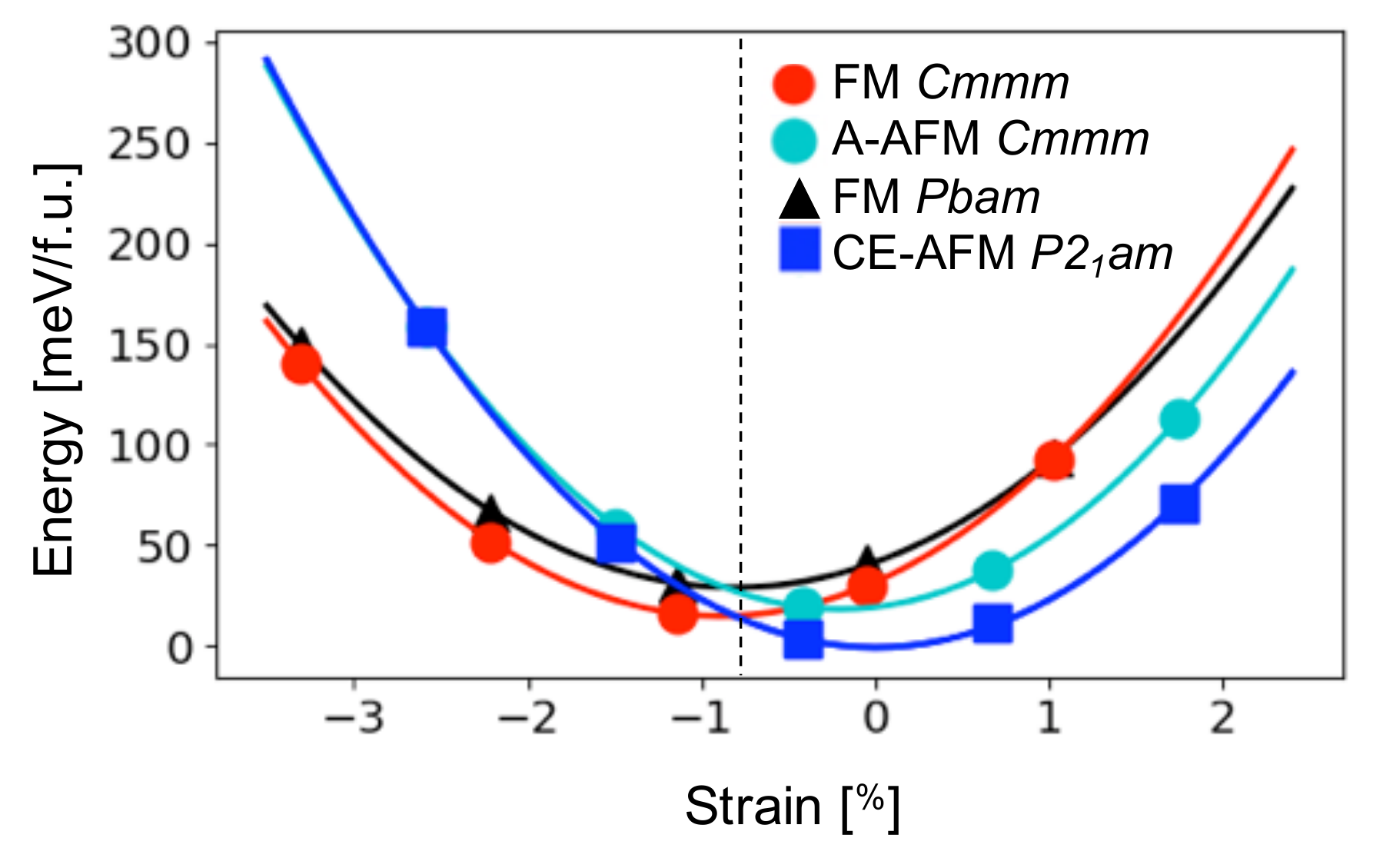}
\caption{\label{fig:strain} { \bf Phase control with epitaxial strain.} Total energy versus strain for several structural and magnetic phases of SmBaMn$_2$O$_6$. The 0$\%$ strain is defined relative to the CE-AFM lattice constants. The dashed line indicates the strain value at which the energy of the CE-AFM $P2_1am$ and FM $Cmmm$ phases are energetically degenerate. Negative and positive strain values correspond to compressive and tensile strains, respectively. }
\end{figure}

Fig.~\ref{fig:strain} shows the  total energy of several structural and magnetic phases of SmBaMn$_2$O$_6$ as a function of epitaxial strain. For tensile and small compressive strains, the polar $P2_1am$ CE-AFM phase is the lowest energy, while at larger compressive strains, the $Cmmm$ FM state becomes the lowest.  At 0.8$\%$ compressive strain, these two phases are energetically degenerate and thus  compete for the ground state. Assuming that the film hosts a mixture of these two states,  this phase competition could enable cross-coupled control of magnetization and polarization: the application of a magnetic field would grow the $Cmmm$ FM regions and shrink the $P2_1am$ polar  regions, so that the magnetization (polarization) increases (decreases). Conversely, the application of an electric field \textcolor{black}{may} increase the  polar regions and decrease the FM regions, so that the polarization (magnetization) increases (decreases). \textcolor{black}{While this is a compelling idea, it is important to note that an applied electric field may instead promote the FM metallic phase in order to screen the field.  Experiments  on Pr$_{0.7}$Ca$_{0.3}$MnO$_3$ have revealed that an electric field of $\approx$7 kV/cm drives an insulator-metal transition.~\cite{Asamitsu1997,Stankiewicz2000}  Coercive fields of various types of ferroelectrics are in the 10-100 kV/cm range, although the field required to change the polarization magnitude (expand the polar regions) rather than completely flip the polarization would be lower. The outcome of electric field application on SmBaMn$_2$O$_6$ would depend on the balance of energy scales of polar domain wall motion and the insulator-metal transition.}

\begin{figure}
\includegraphics[width=0.48\textwidth]{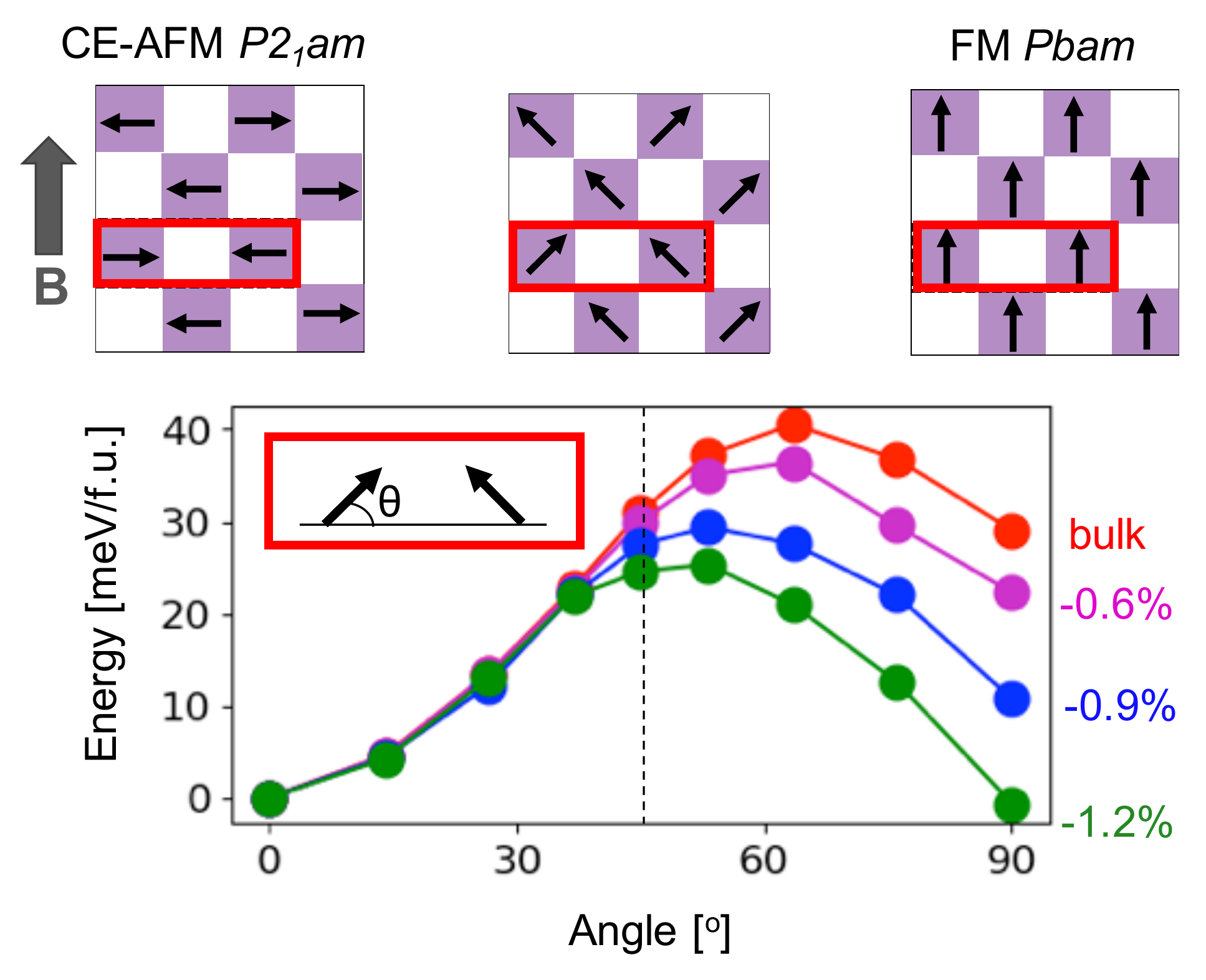}
\caption{\label{fig:cspin} {\bf  Energy barrier between competing phases}. Total energy versus constrained spin angle for bulk and compressively strained SmBaMn$_2$O$_6$.  The angle that the left reference spin (in the red box) makes with the horizontal axis defines the spin angle $\theta$, so  $\theta=0^\circ$ and $90^\circ$ correspond to  CE-AFM and FM, respectively. The dark grey arrow indicates the direction of the hypothetical magnetic field.}
\end{figure}

To understand how strain impacts the barrier for magnetic field control, we consider a $P2_1am$  CE-AFM domain, and imagine applying a magnetic field that is perpendicular to the spin axis as shown in Fig.~\ref{fig:cspin}. This magnetic field  rotates the spins so that they transform to a FM configuration. To describe this transformation, we select a reference spin and define $\theta$ to be the angle that this reference spin makes with the  initial spin axis (so  $\theta=0^\circ$  is CE-AFM,  $\theta=90^\circ$ is FM). Then, for several intermediate $\theta$ we  perform constrained spin calculations, allowing the structure to fully relax at each $\theta$. Since we start with $P2_1am$  and relax the structure at each step, the resulting FM phase at $\theta=90^\circ$ has $Pbam$ symmetry, rather than the lower energy $Cmmm$ (so the barriers in Fig.~\ref{fig:cspin} are an upper limit). As the compressive strain increases, the energy barrier between the  $P2_1am$ CE-AFM and  $Pbam$ FM phases decreases, which suggests that strain lowers the field strength needed to control the competing phases. 
 
\section{\label{sec:discussion}Discussion}
We have shown how the interplay of several structural distortions underlies the coupled structural, electronic, and magnetic ground state of SmBaMn$_2$O$_6$. The coupled distortions induce the CO and an electrical polarization via a hybrid improper ferroelectric mechanism. Our approach  allows a systematic exploration of metastable structures, from which we identify several low-energy FM and A-AFM centrosymmetric metallic phases that compete with the  polar insulating CO/OO/CE-AFM ground state. We show that the domain structure of SmBaMn$_2$O$_6$ visualizes  the coupling and competition between these order parameters, via a set of coupled domains connected by 3- and 4-fold domain wall vortices and antivortices. To assess this domain structure  we suggest microwave impedance microscopy, which already has  been reported for the related  compound Pr(Sr$_{1-x}$Ca$_x$)$_2$Mn$_2$O$_7$~\cite{Ma2015}. 

In addition, we consider two examples where control of the complex SmBaMn$_2$O$_6$ crystal structure provides a knob to manipulate the electronic and magnetic states: at structural domain walls (which are a naturally occurring form of structural modulation in bulk systems), and in epitaxial thin films. In the latter example, we show how it is possible to tune SmBaMn$_2$O$_6$ to a regime where a centrosymmetric FM metallic state is energetically degenerate with the polar insulating bulk ground state, which may enable cross-coupled control of polar and magnetic states.  

\textcolor{black}{Finally, we comment on the applicability of our results to other rare-earth manganites. While we focus here on the special case of $A$-site ordered SmBaMn$_2$O$_6$, the structural distortions as well as the CO/OO/CE-AFM, FM, and A-AFM phases that we investigate occur in many other manganites. In particular, we anticipate that the stripe distortion explored in detail here to be relevant more broadly for half-doped manganites (including $A$-site disordered materials). For example, experiments have reported that $A$-site disordered  La$_{0.5}$Ca$_{0.5}$MnO$_3$~\cite{Radaelli1997,Goff2004} and Tb$_{0.5}$Ca$_{0.5}$MnO$_3$ \cite{Blasco1997} in the CO/OO/CE-AFM phase have symmetry $P2_1/m$. Decomposing the experimental La$_{0.5}$Ca$_{0.5}$MnO$_3$ $P2_1/m$ structure from Ref.~\onlinecite{Goff2004} with respect to cubic $Pm\bar{3}m$ (the reference structure for $A$-site disordered systems) reveals the presence of a large amplitude stripe distortion in this material. We hypothesize that the link between CO and the stripe distortion that we identify in SmBaMn$_2$O$_6$ holds generally for half-doped manganites. We also expect that the coupling between CE-AFM and structural  domains, the scenarios for FM and A-AFM domain wall states, and the tunability of the ground state between CE-AFM and FM states with epitaxial strain to be applicable for both $A$-site ordered and disordered half-doped manganites.}

\textcolor{black}{In contrast, other results depend on the $A$-site ordered structure. For example, $A$-site ordering is required to establish the polar crystal structure of SmBaMn$_2$O$_6$. In addition, we expect that the domain structure of $A$-site disordered half-doped manganites to be different from the 3- and 4-fold domain wall vortex structure of $A$-site ordered SmBaMn$_2$O$_6$ discussed here.  This is because $A$-site ordering  reduces the structural order parameter dimensions by establishing a preferential axis in the high-symmetry reference structure.  For example, with $A$-site ordering the octahedral tilt and stripe distortion order parameters are two- and four-dimensional respectively (as discussed above), while with $A$-site disordering, these order parameters are three- and twelve-dimensional, respectively. As a result, the $A$-site disordered manganites have a larger number of structural domains and more possible paths between domains (domain wall types) than the $A$-site ordered manganites. This would result in a complex domain structure which could be analyzed using the  approach presented in this work.  }


\section{Acknowledgments}
We acknowledge support from the US Department of Energy, Office of Basic Energy Sciences, Division of Materials Sciences and Engineering, under Award No. DE-SC0002334. E.A.N. also acknowledges support from the New Jersey Institute of Technology. This research used computational resources supported by the Cornell University Center for Advanced Computing and by Academic and Research Computing Systems at the New Jersey Institute of Technology.

E. A. N. and J. H. contributed equally to this work. 

\appendix
\section{\label{app:methods}Computational details} 

We perform density functional theory calculations using VASP\cite{Kresse93,Kresse99} and the 
PBEsol exchange-correlation functional~\cite{Perdew08}. We use a plane wave basis with an energy cutoff of 500 eV.
We treat the Mn on-site Coulomb interaction using the Liechtenstein formulation of the DFT + $U$ method \cite{Liechtenstein1995}. We set U=4.0 eV and J=1.2 eV;  these values are chosen because they reproduce the bulk $R$BaMn$_2$O$_6$ phase diagram (see Appendix~\ref{app:u}). 
The unit cell in the  $P2_1am$ CE-AFM ground state phase is $2\sqrt{2}a_{0}\times2\sqrt{2}a_{0}\times2c_{0}$, which contains 80 atoms.    We use a 4$\times$4$\times$6
$\Gamma$-centered \emph{k}-point grid to sample Brillouin zone of the $P2_1am$ CE-AFM phase. For other structures and magnetic orders,  we used smaller computational cells for some calculations where we chose the same density of \emph{k}-points that was used for the 80 atom cell. 
We used a force convergence tolerance of  10 meV/\AA.
Biaxial strain was applied by fixing the in-plane lattice constants $a$ and $b$ to a square and  relaxing the out-of-plane lattice constant $c$ and all atomic positions.
For group theoretical analysis we utilized the ISOTROPY software suite\cite{isotropy}. We visualized crystal structures using VESTA.~\cite{vesta} 


\section{\label{app:u} Choice of $U$ and $J$ parameters}

\begin{figure*}
\begin{center}
\includegraphics[width=0.98\textwidth]{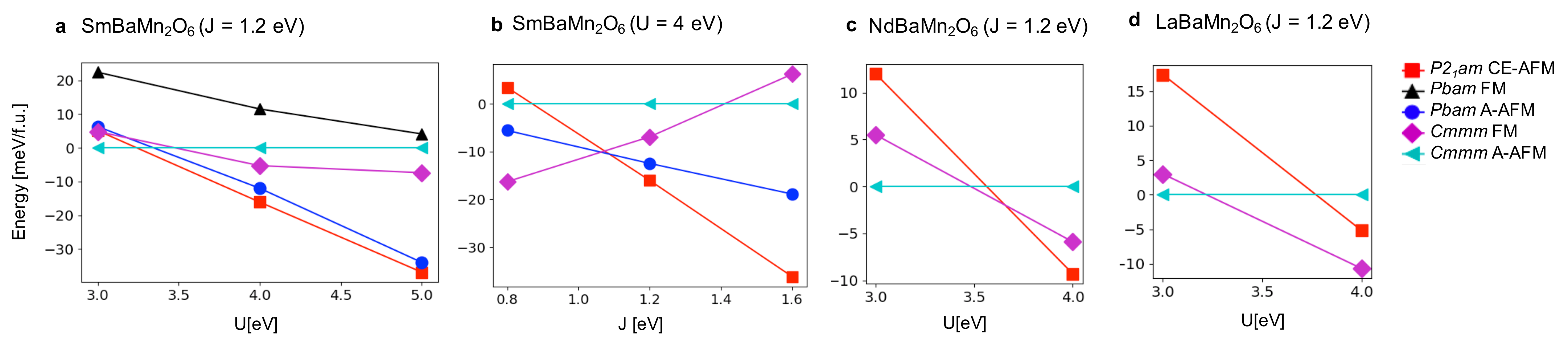}
\caption{\label{fig:vary_U} Total energy of SmBaMn$_2$O$_6$ as a function of (a) $U$ (with fixed $J$=1.2 eV), and (b) $J$ (with fixed $U$=4 eV). Total energy as a function of $U$ for (c) NdBaMn$_2$O$_6$ and (d) LaBaMn$_2$O$_6$. For (c) and (d),  $J$=1.2 eV. All energies are referenced to the $Cmmm$ A-AFM energy.}
\end{center}
\end{figure*}

Table~\ref{tab:phase_diag} summarizes the experimentally reported $R$BaMn$_2$O$_6$ phase diagram, also shown in Fig.~\ref{fig:phase_diag}. The ground state evolves from an AFM CO/OO polar insulator for $R$=Sm, to a A-AFM metal for $R$=Nd and Pr, to a FM metal for $R$=La. However, note that as indicated in Table~\ref{tab:phase_diag}, the experimental picture of the ground state for some systems remains unclear. To guide our choice of $U$ and $J$ parameters in the main text, we explore the bulk structural energetics as we vary these parameters for $R$ = Sm, Nd, and La (Fig.~\ref{fig:vary_U}). First fixing $J$ = 1.2 eV  and varying $U$, we find that for all three compounds (Fig.~\ref{fig:vary_U}a, c, d), the A-AFM state is stabilized for  $U$=3 eV, where the structure has symmetry $Cmmm$ for $R$ = Sm and Nd and $P4/mmm$ for $R$=La. As $U$ increases, the FM state becomes lower energy than the A-AFM state for all three compounds, however, for $R$ = Sm and Nd the CE-AFM state stabilizes more quickly and becomes the ground state for $U$=4 eV. Based on these results, we select $U = 4.0$ eV to use in all other calculations within this work. In Fig.~\ref{fig:vary_U})b, we vary $J$ for SmBaMn$_2$O$_6$ (with $U$ fixed to 4 eV).  For small $J$, the $Cmmm$ FM phase is lowest energy, and then the $P2_1am$ CE-AFM phase stabilizes with increasing $J$. Based on this result, we choose $J$ = 1.2 for all calculations. Note that this set of parameters reproduces the CO/OO CE-AFM ground state for $R$=Sm and the $P4/mmm$ FM metallic ground state for $R$=La. This set of parameters also predicts the  CO/OO CE-AFM  insulating state for $R$=Nd, however we note that as shown in Table~\ref{tab:phase_diag}, there are conflicting reports of the ground state for this system.

\begin{table}
\caption{\label{tab:phase_diag} Experimentally reported crystal and magnetic structures for  $R$BaMn$_2$O$_6$ compounds.}

\begin{tabular}{c  c c c c}
\hline
\hline
$R$ & temperature (K) & space group & magnetic order & reference \\
\hline 
Sm & 400 & $Cmmm$ & - & Ref.~\onlinecite{Sagayama2014} \\
Sm & 300 & $Pnam$ & - & Ref.~\onlinecite{Sagayama2014} \\
Sm & 150 & $P2_1am$ & CE-AFM & Ref.~\onlinecite{Sagayama2014} \\
\hline 
Nd & - & $P4/mmm$ & A-AFM & Ref.~\onlinecite{Nakajima2003} \\
Nd & 290 & $P2_1am$ & - & Ref.~\onlinecite{Yamada2017} \\
\hline
Pr & - & $P4/mmm$ & A-AFM/CE-AFM  & Ref.~\onlinecite{Nakajima2003} \\
& & & coexistence & \\
\hline
La & - & $P4/mmm$ & FM/CE-AFM & Ref.~\onlinecite{Nakajima2003} \\
& & & coexistence & \\
\end{tabular}
\end{table}

\section{\label{app:op}Order parameter notation}

With the exception of the $M_4^+$ breathing distortion, the other structural distortions in SmBaMn$_2$O$_6$ are described by multidimensional order parameters (the $M_5^-$ tilt and $\Gamma_5^-$ polar order parameters are two dimensional, and the $\Sigma_2$ order parameter is four dimensional). In analyzing these multidimensional order parameters, it is important to distinguish between the \textit{amplitude} of the order parameter and its \textit{direction} within the order parameter space. This section describes our notation for these quantities. 

For concreteness, here we consider the two-dimensional $M_5^-$ tilt order parameter ($T_1$, $T_2$). The variables $T_1$ and $T_2$ define the two-dimensional order parameter space, and at any particular point in this space the order parameter amplitude is ${Q}_T = \sqrt{T_1^2+T_2^2}$.  To define the order parameter directions,  we use $a$ and $b$ to denote the  order parameter components $T_i$, where $a$ and $b$ are arbitrary numbers that are not equal to each other. Using this notation, the three distinct directions of the order parameter are: ($T_1$,$T_2$) = ($a$,0), ($a$,$a$), and ($a$,$b$). The key point is that each of these distinct directions defines a family of structures of a particular symmetry (with the order parameter amplitude still a variable). The order parameter amplitude $\tilde{Q}$ that minimizes the total energy for a particular structural symmetry  can be obtained by performing DFT structural relaxations constrained to that symmetry. Finally, for a given structural symmetry (defined by  \textit{distinct} direction of the order parameter) the structural domains are defined by the multiple order parameter directions that are consistent with that symmetry. For example, the $Pmam$ structure, defined by the ($a$,0) distinct direction, has four domains, labelled by the order parameter directions ($a$,0), (-$a$,0), (0,$a$), and (0,-$a$).  

Similarly, the $\Gamma_5^-$ order parameter defines a two dimensional order parameter space ($P_1$,$P_2$), and the $\Sigma_2$ order parameter defines a  four dimensional space ($s_1$,$s_2$,$s_3$,$s_4$).
In the analysis of the domain structure and switching paths in the main text, we label the domains of the $P2_1am$ ground state by the directions of the $\Sigma_2$ order parameter, for example ($s_1$,$s_2$,$s_3$,$s_4$) = ($a$,$b$,0,0), ($a$,-$b$,0,0), etc. It is important to keep in mind that  $a$ and $b$ are arbitrary numbers (with the symmetry constraint that they are unequal), the particular distortion amplitudes realized in the ground state structure ($\tilde{s}_1$, $\tilde{s}_2$) are obtained from structural relaxations. 
We also label the barrier structures by the $\Sigma_2$ order parameter direction that defines their symmetry (where again, the order parameter amplitudes at the barrier are obtained by structural relaxations with the symmetry constrained to that of the barrier).

\section{\label{app:stripe}Stripe distortion}

\begin{figure}
\includegraphics[width=0.48\textwidth]{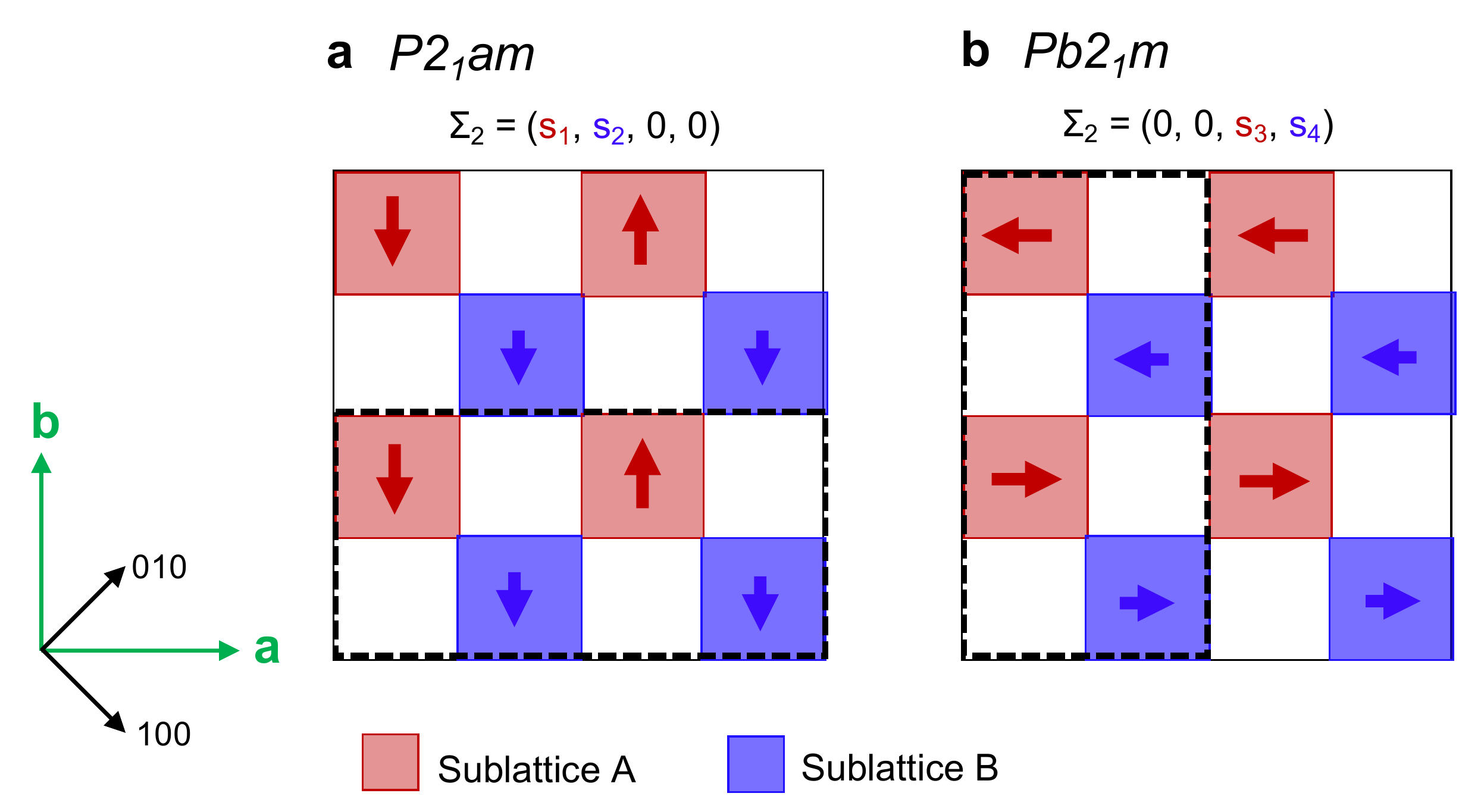}
\caption{\label{fig:sm2-def} {\bf Relationship between sublattices, the stripe distortion, and orthorhombic twins}. Here red and blue represent sublattices A and B, respectively. The red (blue) arrows represent the apical oxygen displacements on  sublattice A (B), all other displacements that contribute to the stripe distortion are suppressed for clarity. With the setting of the orthorhombic axes $a$ and $b$ relative to the tetragonal axes shown, the two orthorhombic twins can be distinguished via different settings of the space group (a) $P2_1am$ and (b) $Pb2_1m$. The distortion forms stripes of upward and downward displacements perpendicular to the long axis, so in twin $P2_1am$ the stripes lie along $b$ and in twin $Pb2_1m$ they lie along $a$. The black dashed line indicates the crystallographic unit cell.}
\end{figure}

The $\Sigma_2$ stripe distortion is most naturally understood by dividing  the MnO$_2$ plane into two sublattices (the same checkerboard sublattices as defined by the CO). Since in principle one can think about sublattices without CO, we call these sublattices A and B in the following discussion for generality. The relationship between these sublattices, the stripe distortion, and the orthorhombic twins is shown in Fig.~\ref{fig:sm2-def}. 

Once the setting of the orthorhombic axes $a$ and $b$ is defined relative to the tetragonal axes as shown in Fig.~\ref{fig:sm2-def}, the two orthorhombic twins are labelled by different space group settings, $P2_1am$ and $Pb2_1m$, shown in Fig.~\ref{fig:sm2-def}(a) and (b), respectively. 
The first two components of the four dimensional $\Sigma_2$ order parameter ($s_1$ and $s_2$) give the stripe distortion amplitude for the domains in the $P2_1am$ orthorhombic twin ($s_3$=$s_4$=0). Here $s_1$ gives the distortion amplitude on sublattice A, and $s_2$ gives the distortion amplitude on sublattice B; the apical oxygen displacements on the two sublattices are indicated by red (blue) arrows  (the rest of the displacements that contribute to the stripe distortion are suppressed for clarity). The last two components of the $\Sigma_2$ order parameter ($s_3$ and $s_4$) give the distortion amplitude for the domains in the $Pb2_1m$ orthorhombic twin ($s_1$=$s_2$=0), with $s_3$ ($s_4$) giving the amplitude on sublattice A (B). In the $P2_1am$ twin, there are stripes of upward and downward displacements parallel to the $b$ axis, while in the $Pb2_1m$ twin, the stripes lie along $a$.  Finally, if one only considers domains within one orthorhombic twin, the $\Sigma_2$ order parameter can be treated as a two-dimensional quantity ($s_1$,$s_2$) as is done in the first part of the main text for simplicity. 

\begin{figure*}
\begin{center}
\includegraphics[width=0.9\textwidth]{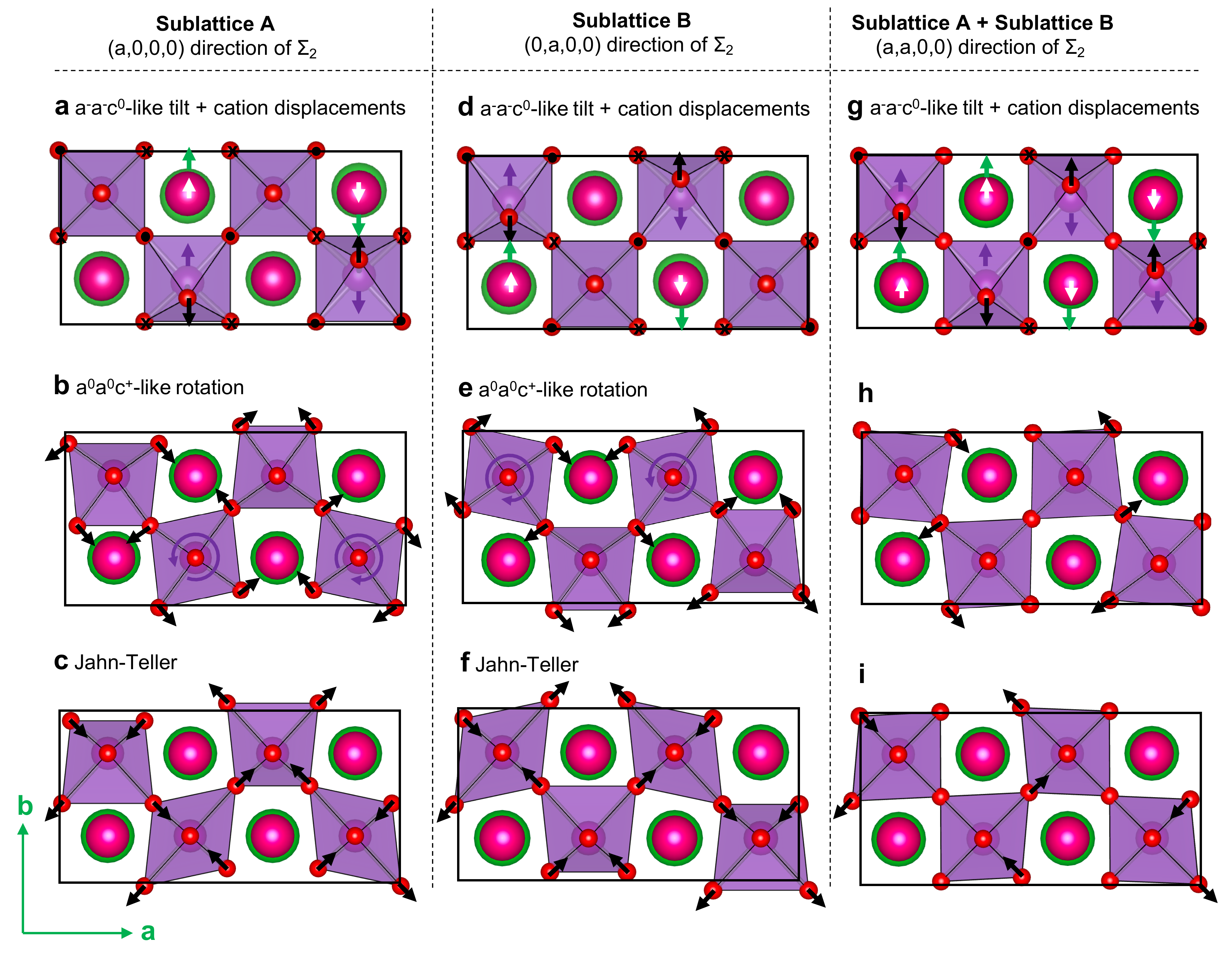}
\caption{\label{fig:sm2-supp} {\bf Atomic displacements that contribute to the stripe distorion}.  The left column shows distortions present on sublattice A only: (a) $a^-a^-c^0$ octahedral tilt-like distortion and cation displacements, (b) $a^0a^0c^+$ octahedral rotation-like distortion, and (c) Jahn-Teller distortion. These distortions transform like the $(a,0,0,0)$ direction of $\Sigma_2$ and establish a structure with symmetry $Pmam$. The middle column shows distortions on sublattice B only: (d) $a^-a^-c^0$ octahedral tilt-like distortion and cation displacements, (e) $a^0a^0c^+$ octahedral rotation-like distortion, and (f) Jahn-Teller distortion, which transform like the $(0,a,0,0)$ direction of $\Sigma_2$ and establish a different $Pmam$ domain. The right column (g-i) shows the case where equal amplitudes of these distortions are present on both sublattices, which yields structures that transform like  the $(a,a,0,0)$ direction of $\Sigma_2$ (symmetry $Pbam$). If the distortion amplitudes on the two sublattices are different (the ($a$,$b$,0,0) direction of $\Sigma_2$), then the $P2_1am$ structure is established. Displacements of the oxygen atoms are indicated by black arrows, and the Mn, Sm, and Ba displacements are shown with purple, white, and green arrows, respectively. The amplitudes of all distortions are artificially increased for clarity.}
\end{center}
\end{figure*}

Dividing the  complex set of atomic displacements that contribute to the stripe distortion into physically intuitive groups helps us organize our thinking about this distortion. These physically intuitive displacements are shown in Fig.~\ref{fig:sm2-supp}(a-c) on 
 sublattice A and (d-f) on sublattice B. 
 Separate panels show the different  displacements: the $a^-a^-c^0$ octahedral tilt-like distortion and cation displacements (panels a and d), the $a^0a^0c^+$ octahedral rotation-like distortion (panels b and e) and the Jahn-Teller distortion (panels c and f). If displacements are only allowed on one sublattice at a time as in Fig.~\ref{fig:sm2-supp}(a-c) and (d-f), the symmetry of the structure is $Pmam$, defined by the ($a$,0,0,0) direction of $\Sigma_2$ (note that this is a distinct structure from the $Pmam$ defined by the ($a$,0) direction of $M_5^-$). If displacements are allowed on both sublattices, but have equal amplitude ($s_1=s_2$) the symmetry is $Pbam$ (shown in Fig.~\ref{fig:sm2-supp}(g-i)), and if they are of different amplitude ($s_1 \ne s_2$) the symmetry is $P2_1am$.
 Note that the displacement amplitudes in Fig.~\ref{fig:sm2-supp} have been artificially increased for clarity, so the displacement patterns shown should be viewed as the symmetry-allowed displacements, not the actual patterns obtained in relaxed structures.. The relative contributions of the different displacement types to the total stripe distortion amplitude (as well as the amplitude itself) are determined from DFT+$U$ structural relaxations. 
 
 To make this point clear, Table~\ref{tab:sm2-more} shows the decomposition of the stripe distortion into physically intuitive displacements, obtained from DFT+$U$ structural relaxations with the symmetry constrained to several isotropy subgroups of $\Sigma_2$. 
It is clear from Table~\ref{tab:sm2-more} that the total distortion amplitude as will as the relative contribution of the different displacement types  depends on the magnetic order.  
Starting with the $Pmam$ structure with FM order, the largest contribution to the stripe distortion is the apical oxygen displacements in the SmO layer, followed by the $O_{eq}$ and Sm displacements (thus, the displacement pattern in the relaxed structure looks most similar to Fig.~\ref{fig:sm2-supp}(a), with the symmetry-allowed displacements in (b) and (c) having negligible amplitudes). In contrast, the stripe distortion in the $Pmam$ structure with A-AFM order has a much larger amplitude, and the relative contributions of the displacement types are quite different, in particular the $a^0a^0c^+$ rotation-like displacement now makes a significiant contribution, and the Jahn-Teller distortion is also present. 
The remaining columns of  Table~\ref{tab:sm2-more} show the decomposition of the stripe distortion for structures with $Pbam$ and $P2_1am$ symmetry and different magnetic orders. There are variations in the total amplitude as well as the relative contributions of the different displacement types.  However, for all structural symmetries and magnetic orders,  the $a^-a^-c^0$ tilt-like displacements and the cation displacements  are present, while the $a^0a^0c^+$ rotation-like displacement and Jahn-Teller distortion only contribute in certain cases.

\begin{table*}
\begin{center}
\caption{\label{tab:sm2-more} Decomposition of the $\Sigma_2$ stripe distortion into physically intuitive displacements (coefficients $A_{i\Sigma_2}$ obtained from a structural decomposition) on sublattices A and B, obtained from DFT structural relaxations with several symmetries and magnetic orders. The decomposition of the stripe distortion obtained from the $P2_1am$ CE-AFM structure, already presented in Table~\ref{tab:sm2} is reproduced here for ease of comparison. For phases with CO, sublattice A (B) corresponds to the Mn$^{3+}$ (Mn$^{4+}$) sublattice. The amplitudes are reported in the 40 atom crystallographic unit cell in \AA. }

\begin{tabular}{c |c | c   c | c  c | c c | c c | c c | c c}
\hline
\hline
$i$ & Distortion  &  \multicolumn{2}{c}{$Pmam$ FM} & \multicolumn{2}{c}{$Pmam$ A-AFM} &  \multicolumn{2}{c}{$Pbam$ FM} &  \multicolumn{2}{c}{$Pbam$ A-AFM} &  \multicolumn{2}{c}{$P2_1am$ A-AFM} &  \multicolumn{2}{c}{$P2_1am$ CE-AFM} \\
 & &  $\tilde{s}_\mathrm{A}$ & $\tilde{s}_\mathrm{B}$ &  $\tilde{s}_\mathrm{A}$ & $\tilde{s}_\mathrm{B}$ & $\tilde{s}_\mathrm{A}$ & $\tilde{s}_\mathrm{B}$ &  $\tilde{s}_\mathrm{A}$ & $\tilde{s}_\mathrm{B}$ & $\tilde{s}_\mathrm{A}$ & $\tilde{s}_\mathrm{B}$ & $\tilde{s}_\mathrm{A}$ & $\tilde{s}_\mathrm{B}$\\
\hline
1 & Jahn-Teller (O$_{eq}$) &  0.0 & 0.0 & 0.11 & 0.0 & 0.01 & 0.01 & 0.02 & 0.02 & 0.16 & 0.03 & 0.24 & 0.02\\
\hline
2 & $a^0a^0c^+$  (O$_{eq}$) &0.04 & 0.0 & 0.27 & 0.0 & 0.06 & 0.06 & 0.15 & 0.15 & 0.33 & 0.05 &0.46 & 0.06\\
\hline
& $a^-a^-c^0$ & & & & \\
3 & O$_{ap}$ (Sm layer) & 0.16 & 0.0 & 0.36 & 0.0 & 0.23 & 0.23 & 0.30 & 0.30 & 0.43 & 0.25 &  0.53 & 0.26\\
4 & O$_{ap}$ (Ba layer) & 0.02 & 0.0 & 0.0 & 0.0 & 0.03 & 0.03 & 0.04 & 0.04 & 0.02 & 0.07 &  0.05 & 0.08\\
5 & O$_{eq}$ & 0.08 & 0.0 & 0.18 & 0.0 &  0.14 & 0.14 & 0.18 & 0.18 & 0.23 & 0.19 & 0.28 & 0.21\\
\hline
6 & Mn & 0.01 & 0.0 & 0.06 & 0.0 &   0.01 & 0.01 &0.02 & 0.02 & 0.11 & 0.00 &  0.20 & 0.00\\
7 & Sm & 0.09 & 0.0 & 0.06 & 0.0 &   0.08 & 0.08 &0.08 & 0.08 & 0.01 & 0.09 &  0.05 & 0.07\\
8 & Ba & 0.01 & 0.0 & 0.04 & 0.0 &  0.01 & 0.01 &0.02 & 0.02 & 0.09 & 0.01 &  0.13 & 0.01\\
\hline
 & total &  0.21 & 0.0& 0.50 & 0.0 &  0.37 & 0.37 & 0.40 & 0.40 & 0.63 & 0.34 &  0.83 & 0.36\\

\end{tabular}
\end{center}
\end{table*}

Finally,  Table~\ref{tab:sm2-more} shows that the displacements on the Mn$^{3+}$ sublattice (sublattice A) are much larger than those on the Mn$^{4+}$ sublattice for both the $P2_1am$ A-AFM and CE-AFM phases. By definition, the Jahn-Teller distortion is only present on the Mn$^{3+}$ sublattice, but why are the octahedral tilt- and rotation-like displacements also larger?
We can answer this question using the following  crystal chemical argument. Once there is some amount of CO (Mn$^{3.5+\delta}$/Mn$^{3.5-\delta}$) the Mn ions are different sizes: the Shannon radii of Mn$^{3+}$ and Mn$^{4+}$ in octahedral coordination are 0.645 and 0.53, respectively. Recalling that the tolerance factor for an ABO$_3$ perovskite is
$\tau = ({r_A+r_O})/({\sqrt{2}(r_B+r_O)})$, 
we can then think about separate tolerance factors for the Mn$^{3+}$ and Mn$^{4+}$ sublattices. Since $r_\mathrm{Mn3+} > r_\mathrm{Mn4+}$, then $\tau_\mathrm{Mn3+} < \tau_\mathrm{Mn4+}$, rationalizing the larger amplitude of octahedral tilt- and rotation-like distortions on the Mn$^{3+}$ sublattice. 

\section{\label{app:decomp}Structural decomposition and energy surface calculations}

This section provides more details on our methods for structural decomposition and energy surface calculations. These are related analyses, because in a structural decomposition we take a distorted structure (either from DFT  or from experiment) and decompose it into symmetry adapted modes that transform like irreps of the high symmetry reference structure $P4/mmm$, while for the energy surface calculations we start with $P4/mmm$ and freeze in symmetry adapted modes to create distorted structures.

Let ${\bf R}_{P4/mmm}$ and ${\bf R}_{LS}$ be  vectors containing the atomic positions for the high-symmetry reference structure $P4/mmm$ and a low-symmetry (LS) structure (which could be $P2_1am$, $Cmmm$, etc). We can then write
\begin{equation}
{\bf R}_{LS} = {\bf R}_{P4/mmm} + {\bf u}
\end{equation}
where ${\bf u}$ is a vector containing displacements of the atoms from their high-symmetry positions. To perform a symmetry analysis of the displacements,  ${\bf u}$ can  be decomposed into symmetry adapted modes that transform like irreps of $P4/mmm$. For the $P2_1am$ ground state structure ${\bf u}$ can be written as:
\begin{eqnarray}
{\bf u} &=& \sum_{i=1}^2A_{i\Gamma_1^+}{\bf e}_{i\Gamma_1^+} + \sum_{i=1}^7 A_{i\Gamma_5^-}{\bf e}_{i\Gamma_5^-}+ \sum_{i=1}^6 A_{iM_5^-}{\bf e}_{iM_5^-} \\ \nonumber
&+& \sum_{i=1}^2 A_{iM_4^+}{\bf e}_{iM_4^+} + \sum_{i=1}^1 A_{iM_1^+}{\bf e}_{iM_1^+}+ \sum_{i=1}^{16} A_{i\Sigma_2}{\bf e}_{i\Sigma_2}.
\end{eqnarray}
Here ${\bf e}_{i\tau}$ labels a normalized symmetry adapted mode that transforms like the irrep $\tau$ of $P4/mmm$, where $\tau = \{ \Gamma_1^+, \Gamma_5^-, M_5^-, M_4^+, M_1^+, \Sigma_2 \}$. The index $i$ sums over the number of modes that transform like each irrep, where the total number of modes is equal to the total number of free atomic positions in the structure (for $P2_1am$, there are 33). Note that we do not discuss the $\Gamma_1^+$ and $M_1^+$ modes in the main text because $\Gamma_1^+$ describes strain modes that maintain the $P4/mmm$ symmetry, and the $M_1^+$ mode has zero amplitude in all structures that we consider. The coefficient $A_{i\tau} = {\bf u} \cdot {\bf e}_{i\tau}$ gives the amplitude that mode ${\bf e}_{i\tau}$ contributes to the total distortion amplitude $|{\bf u}|$. The amplitudes reported in Table~\ref{tab:decomp} are obtained by summing over all modes $i$ that transform like a given irrep $\tau$, that is: $A_\tau = \sqrt{\sum_iA_{i\tau}^2}.$

To calculate the energy surfaces presented in Fig.~\ref{fig:surf}, we start with the high-symmetry reference structure $P4/mmm$ and freeze in increasing amplitudes of the symmetry adapted modes described above. As a concrete example, to calculate the $M_5^-$ energy surface shown in the left panel of Fig.~\ref{fig:surf}, we construct $j$ structures (with symmetry $Pmam$)
\begin{equation}
{\bf R}_{Pmam, j} = {\bf R}_{P4/mmm} + B_j {\bf e}_{M_5-},
\end{equation}
where $B_j$ is the mode amplitude frozen into the $j^{th}$ structure,  
\begin{equation}
\label{eqn:mode}
{\bf e}_{M_5^-} = \frac{1}{N_{M_5^-}}\sum_{i=1}^6A_{iM_5^-}{\bf e}_{iM_5^-}
\end{equation}
is the normalized sum of all modes that transform like $M_5^-$, and 
\begin{equation}
N_{M_5^-} = \sqrt{\sum_{i=1}^6A_{iM_5^-}^2}
\end{equation}
 is the normalization coefficient.  From Eq.~\ref{eqn:mode}, it is clear that to construct ${\bf e}_{M_5^-}$, values for the $A_{iM_5^-}$ coefficients are needed. We obtain these from structural decompositions of the DFT+$U$-relaxed structures. Table~\ref{tab:m5} shows the coefficients $A_{iM_5^-}$ obtained from structural decompositions of DFT+$U$-relaxed structures with $Pmam$, $Cmmm$, and $P2_1am$ symmetry. There are slight variations between the relative contributions of the different $A_{iM_5^-}$ depending on the structural symmetry. However, the differences are small enough that the resulting $M_5^-$ energy surface does not depend much on which $A_{iM_5^-}$ we use in the calculation. We choose to use the $A_{iM_5^-}$ in Table~\ref{tab:m5} obtained from the $P2_1am$ A-AFM structural relaxation (the reason for this choice will become clear below).
 
We can write down analogous expressions for the symmetry adapted modes that use use to compute the other energy surfaces. The question of which coefficients $A_{i\Sigma_2}$ to use in the calculation of the $\Sigma_2$ energy surface is more complicated for two reasons, both of which are evident from Table~\ref{tab:sm2-more}. Note that there are 16 such coefficients, 8 coming from the displacements on each sublattice (see Table~\ref{tab:sm2-more}). First, the relative contributions of the various $A_{i\Sigma_2}$ depend strongly on the structural symmetry, for example the modes that correspond to the Jahn-Teller and $a^0a^0c^+$ rotation-like distortion only contribute in certain structures. Second, even within the same structural symmetry, the relative sizes of the $A_{i\Sigma_2}$ coefficients depend on the magnetic order imposed in the structural relaxation.   Thus we must choose which set of $A_{i\Sigma_2}$ from Table~\ref{tab:sm2-more} to use in the construction of ${\bf e}_{\Sigma_2}$, and perform the energy surface calculations with the same magnetic order as was used in the calculation of the $A_{i\Sigma_2}$ coefficients. 
 
 Since the primary goal of our energy surface calculations is to understand the instabilities of $P4/mmm$ as well as the coupling terms that stabilize the ground state, we choose to construct  ${\bf e}_{\Sigma_2}$ using coefficients $A_{i\Sigma_2}$  from a $P2_1am$ structural relaxation, because this is the form of the distortion that couples to CO. As is evident from Table~\ref{tab:sm2-more}, the $A_{i\Sigma_2}$ obtained from relaxation with $P2_1am$ symmetry and A-AFM and CE-AFM orders are similar (but not identical). We choose to use the A-AFM coefficients (and calculate all energy surfaces in Fig.~\ref{fig:surf} with A-AFM)  because this allows us to focus on the energetics of the coupled structural distortions, without the additional energy contribution from the coupling between CO and CE-AFM. However, our conclusions do not depend on this choice, see the Supplementary Information~\cite{foot1} for energy surfaces calculated using other choices of $A_{i\Sigma_2}$.

\begin{table}
\caption{\label{tab:m5} Coefficients $A_{iM_5^-}$ obtained from structural decomposition of several structures of different symmetry that contain the $M_5^-$ distortion. Note that for $Pmam$ and $Cmmm$, the coefficients and the same regardless of whether FM or A-AFM order is imposed, while for $P2_1am$ there are small differences between the results with CE-AFM and A-AFM orders (listed separately).  The coefficients are in units of \AA, and are given for a 20 atom cell for $Pmam$ and $Cmmm$ and a 40 atom cell for $P2_1am$.  }

\begin{tabular}{c |c | c   | c | c | c}
\hline
\hline
 $i$ & displacement  &  $Pmam$ & $Cmmm$ & $P2_1am$  & $P2_1am$ \\
 & & & & CE-AFM & A-AFM \\
\hline
1 & O$_{ap}$ (Sm layer) & 0.36  & 0.46 & 0.57 & 0.55\\
 2 &O$_{ap}$ (Ba layer) & 0.15  & 0.17 & 0.24 & 0.23\\
 3 & O$_{eq}$ & 0.34 & 0.40 & 0.56 & 0.53\\
\hline
 4 & Mn & 0.01 & 0.01 & 0.02 & 0.01\\
 5 & Sm & 0.09 & 0.12 & 0.11 & 0.12\\
 6 &Ba & 0.03 & 0.03 & 0.02 & 0.02\\
\hline
$N_{M_5^-}$ & total &  0.53  & 0.65 & 0.84 & 0.80\\

\end{tabular}
\end{table}

\section{\label{app:energy}Free energy expansion and domains}

For simplicity, the free energy expansions in the main text are given for one orthorhombic twin domain. With the full multidimensional order parameters, which are required to treat both twins at the same time, the coupling terms between structural order parameters in Eqs.~\ref{eq:tss}-~\ref{eq:tcp} are:

\begin{equation}
F_{tss} = \delta_{tss}(T_1s_3s_4 - T_2s_1s_2),
\end{equation}
\begin{equation}
F_{bss} = \delta_{bss}Q_b(s_1^2-s_2^2 + s_3^2-s_4^2),
\end{equation}
and
\begin{equation}
F_{tbp} = \delta_{tbp}[T_1Q_b(P_1-P_2) + T_2Q_b(P_1+P_2)].
\end{equation}

These free energy expansions are used to derive the order parameter directions of the $P2_1am$ structural domains shown in Table~\ref{tab:domains}.

\begin{figure}
\includegraphics[width=0.49\textwidth]{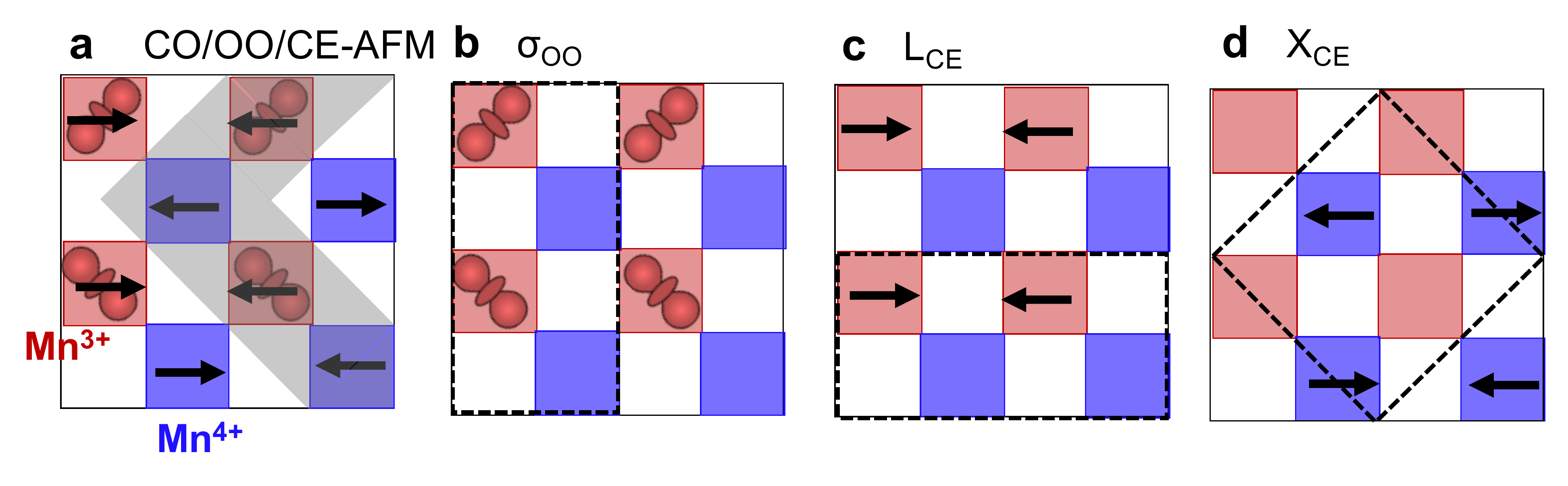}
\caption{\label{fig:ce} {\bf CO, OO, and CE-AFM order parameters}. (a) The CO/OO/CE-AFM phase with Mn$^{3+}$ and Mn$^{4+}$ sites  colored red and blue, respectively.  The order parameters are shown individually for (b) OO, (c) magnetic order $L_\mathrm{CE}$ on the Mn$^{3+}$ sublattice, and (d) magnetic order $X_\mathrm{CE}$ on the Mn$^{4+}$ sublattice. The black dashed lines indicate the unit cell for each order parameter. }
\end{figure}

Fig.~\ref{fig:ce}a summarizes the relationship between the CO, OO, and CE-AFM order parameters.  The $d_{x^2-r^2}/d_{y^2-r^2}$ OO (Fig.~\ref{fig:ce}b), which we denote by the order parameter $\sigma_{OO}=(\sigma_1,\sigma_2,\sigma_3,\sigma_4)$,  transforms like the $(a,0,0,0)$ direction of $\Sigma_2$ and thus couples bilinearly to the stripe distortion. 
 The two magnetic order parameters that describe the CE-AFM state  are shown in Fig.~\ref{fig:ce}c,d. The first, $L_\mathrm{CE}=(L_1,L_2,L_3,L_4)$, transforms like the (0,0,$a$,0) direction of the $m\Sigma_1$ irrep and describes the magnetic ordering on the Mn$^{3+}$ sublattice, while the second, $X_\mathrm{CE}=(X_1,X_2)$,  transforms like the ($a$,0) direction of the $mX_1^+ - mX_2^+$ irrep and describes the magnetic ordering on the Mn$^{4+}$ sublattice (the $m$ indicates  \textit{magnetic} irreps, where time reversal changes the sign of all magnetic moments~\cite{isotropy}). The magnetic point group established by  CE-AFM with spins lying in the $ab$ plane is $P_bmc2_1$.

Note that in the main text we treated $L_{CE}$ and $X_{CE}$ as two- and one-dimensional order parameters, respectively, which restricted that analysis to one orthorhombic twin. With the full order parameters presented here, the free energy expansion given in Eq.~\ref{eq:ce} then becomes: 
\begin{eqnarray}
\label{eq:f_ce}
F_{cL}&=& \eta_{cL} Q_{CO} (L_1^2-L_2^2+L_3^2-L_4^2)    \\
&+& \eta_{cX}Q_{CO}(X_1^2-X_2^2) \nonumber \\
 &+& \eta_{\sigma LX}(\sigma_1L_3X_1 + \sigma_3L_1X_1 + \sigma_2L_4X_2+\sigma_4L_2X_2) \nonumber \\
  &+&  \eta_{LX}(L_1^2-L_2^2+L_3^2-L_4^2)(X_1^2-X_2^2) \nonumber
\end{eqnarray}
This expansion is used to determine the OO and CE-AFM domains that couple to the structural domains of $P2_1am$, see Table~\ref{tab:domains}.

\begin{table*}
\begin{center}
\caption{\label{tab:domains} Coupled structural, OO, and CE-AFM domains in the $P2_1am$ CE-AFM ground state of SmBaMn$_2$O$_6$. There are 16 structural domains, and 32 magnetic domains, so each structural domain couples to two magnetic domains. The other 16 magnetic domains can be reached from those shown here by flipping the signs on both magnetic order parameters. Domains in the two orthorhombic twins are distinguished by the space group setting ($P2_1am$ and $Pb2_1m$).}

\begin{tabular}{c | c | c | c | c | c | c | c | c}
\hline
\hline
Domain & $M_5^-$ & $M_4^+$  &$\Sigma_2$ & $\Gamma_5^-$ & space group & OO & ${\bf L}_\mathrm{CE}$ & ${\bf X}_\mathrm{CE}$\\
\hline
1 & (0,$T$) & $-Q_{b}$ & ($b$, -$a$, 0, 0) & (-$P$,-$P$) & $P2_1am$ & (0,-$\sigma$,0,0) & (0,0,0,$L$) & (0,-$X$)\\
2 & (0,$T$) & $-Q_{b}$ & (-$b$, $a$, 0 , 0) & (-$P$,-$P$) & $P2_1am$& (0,$\sigma$,0,0)& (0,0,$L$) & (0,$X$)\\
3 & (0,$T$) & $Q_{b}$ & ($a$, -$b$, 0, 0) & ($P$,$P$) & $P2_1am$& ($\sigma$,0,0,0) & (0,0,$L$,0) & ($X$,0)\\
4 & (0,$T$) & $Q_{b}$ & (-$a$, $b$, 0, 0) & ($P$,$P$) & $P2_1am$& (-$\sigma$,0,0,0)& (0,0,$L$,0) & (-$X$,0)\\
\hline
5 & (0,-$T$) & $-Q_{b}$ & ($b$, $a$, 0, 0) & ($P$,$P$) & $P2_1am$ & (0,$\sigma$,0,0) &(0,0,0,$L$) & (0,$X$)\\
6 & (0,-$T$) & $-Q_{b}$ & (-$b$, -$a$, 0, 0) & ($P$,$P$) & $P2_1am$& (0,-$\sigma$,0,0) &(0,0,0,$L$) & (0,-$X$)\\
7 & (0,-$T$) & $Q_{b}$ & ($a$, $b$, 0, 0) & (-$P$,-$P$) & $P2_1am$& ($\sigma$,0,0,0) & (0,0,$L$,0) & ($X$,0)\\
8 & (0,-$T$) & $Q_{b}$ & (-$a$, -$b$, 0, 0) & (-$P$,-$P$) & $P2_1am$& (-$\sigma$,0,0,0)& (0,0,$L$,0) & (-$X$,0)\\
\hline
9 & ($T$,0) & $Q_{b}$ & (0, 0, $a$,$b$) & ($P$,-$P$) & $Pb2_1m$ &(0,0,$\sigma$,0) & ($L$,0,0,0) & ($X$,0)\\
10 & ($T$,0) & $Q_{b}$ & (0, 0,-$a$,-$b$) & ($P$,-$P$) & $Pb2_1m$&(0,0,-$\sigma$,0) & ($L$,0,0,0) & (-$X$,0)\\
11 & ($T$,0) & $-Q_{b}$ & (0, 0, $b$, $a$) & (-$P$,$P$) & $Pb2_1m$&(0,0,0,$\sigma$) & (0,$L$,0,0) & (0,$X$)\\
12 & ($T$,0) & $-Q_{b}$ & (0, 0, -$b$, -$a$) & (-$P$,$P$) & $Pb2_1m$&(0,0,0,-$\sigma$) & (0,$L$,0,0) & (0,-$X$)\\
\hline
13 & (-$T$,0) & $Q_{b}$ & (0, 0, $a$, -$b$) & (-$P$,$P$) & $Pb2_1m$ & (0,0,$\sigma$,0) & ($L$,0,0,0) & ($X$,0)\\
14 & (-$T$,0) & $Q_{b}$ & (0, 0, -$a$, $b$) & (-$P$,$P$) & $Pb2_1m$&(0,0,-$\sigma$,0) & ($L$,0,0,0) & (-$X$,0)\\
15 & (-$T$,0) & -$Q_{b}$ & (0, 0, $b$, -$a$) & ($P$,-$P$) & $Pb2_1m$&(0,0,0,-$\sigma$) & (0,$L$,0,0) & (0,-$X$)\\
16 & (-$T$,0) & -$Q_{b}$ & (0, 0, -$b$, $a$) & ($P$,-$P$) & $Pb2_1m$&(0,0,0,$\sigma$) & (0,$L$,0,0) & (0,$X$)\\

\end{tabular}
\end{center}
\end{table*}

\section{\label{app:phase}Phase control with chemical substitution}

Given that SmBaMn$_2$O$_6$ has a $P2_1am$ CE-AFM ground state, and LaBaMn$_2$O$_6$ is a FM metal, chemical substitution (Sm$_{1-x}$La$_x$BaMn$_2$O$_6$) may be an alternative strategy to tune the material such that these two ground states compete. \textcolor{black}{We note that this strategy has been tried experimentally, and while phase separation into FM and CO insulating regions was observed, no CMR effect was detected}~\cite{Ueda2007}. 

\begin{figure*}
\begin{center}
\includegraphics[width=0.8\textwidth]{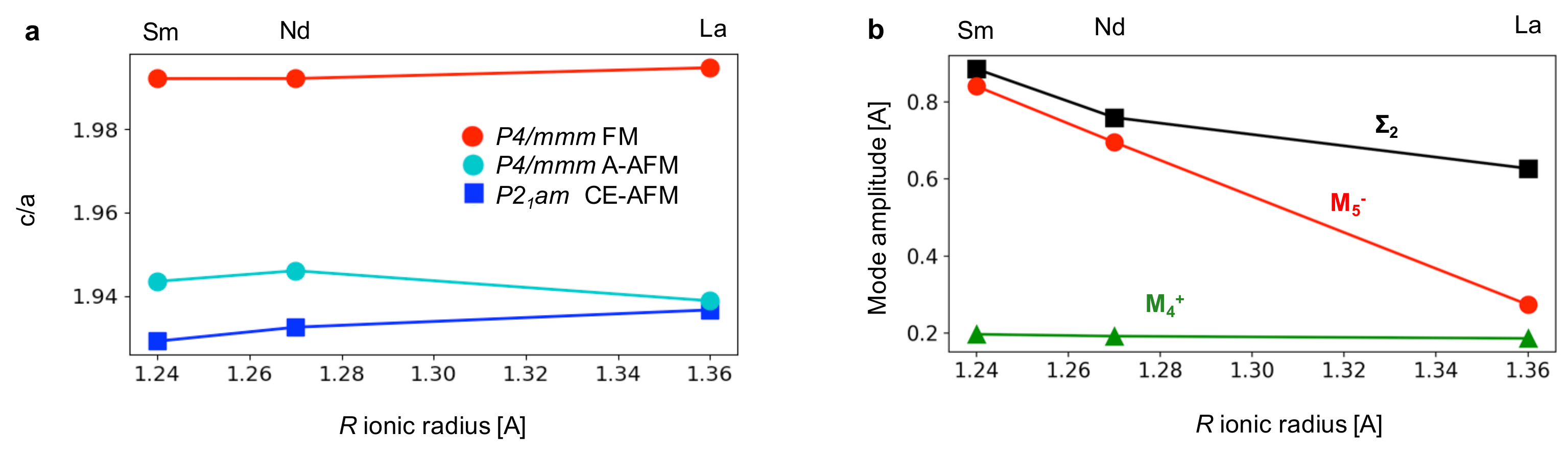}
\caption{\label{fig:doping} { Evolution of the crystal structure of $R$BaMn$_2$O$_6$ with change to the rare  earth $R^{3+}$ ionic radius:} (a)  $c/a$ ratio for the $P2_1am$ CE-AFM and the $P4/mmm$ FM and A-AFM phases, and (b)  the symmetry adapted mode amplitudes for the $P2_1am$ CE-AFM phase. }
\end{center}
\end{figure*}

One of the main effects of chemical substitution is to change the average $R^{3+}$ ionic radius. We can access this effect just by varying the $R$ cation in $R$BaMn$_2$O$_6$, which gives access to the essential physics but is simpler than  including dopants in our calculations.  Fig.~\ref{fig:doping}(a) shows the $c/a$ ratio for a few selected structures with $R$ = Sm, Nd, and La. As discussed in the main text, the $c/a$ ratio is different depending on whether the spins are FM- or AFM- coupled along the $c$ axis: for FM coupling $c/a \approx 1.98$, while for the AFM coupling, $c/a\approx 1.94$. The $c/a$ ratio hardly changes as we vary the ionic radius of $R$. \textcolor{black}{This shows that doping La to form Sm$_{1-x}$La$_x$BaMn$_2$O$_6$ may be a less effective way of tuning the ground state than epitaxial strain, because it does not change the $c/a$ ratio which couples strongly to the magnetic state.  }

Fig.~\ref{fig:doping}(b) shows the symmetry adapted mode amplitudes for the $P2_1am$  CE-AFM phase, again for $R$=Sm, Nd, and La. The most noticeable feature is that the amplitude of the $M_5^-$ octahedral tilt reduces significantly, which reflects that the  tolerance factor is getting closer to 1. The $\Sigma_2$ stripe distortion also decreases somewhat. However, the amplitude of the $M_4^+$ breathing distortion, which directly couples to the magnetic states,  hardly changes. 

\bibliographystyle{apsrev}   
\bibliography{nowadnick-sbmo}

\end{document}